\documentclass{acseEA} 
\usepackage{amsmath}
\usepackage{paralist}
\usepackage[misc]{ifsym}
\usepackage{epsfig} 
\usepackage{epstopdf} 
\usepackage[colorlinks=true]{hyperref}
\hypersetup{urlcolor=blue, citecolor=red}
\allowdisplaybreaks

\def\currentvolume{X}
 \def\currentissue{X}
  \def\currentyear{200X}
   \def\currentmonth{XX}
    \def\ppages{X-XX}
     \def\DOI{10.3934/acse.2023015}

\usepackage{pgf-pie}
\usetikzlibrary{matrix, shapes, arrows, calc, positioning, decorations.pathreplacing, decorations}
\tikzset{lines/.style={draw=none}}
\usepackage{verbatim}
\usepackage{subcaption}
\usepackage{bm}
\usepackage{caption}
\usepackage{subcaption}
\usepackage{mwe}
\usepackage{enumitem}
\usepackage{color}
\usepackage{array}
\usepackage{multirow}
\usepackage{lipsum}
\usepackage{pgfplots}
\usepackage[toc,page]{appendix}
\usepackage{epstopdf}
\usepackage{longtable}
\usepackage{mathtools}
\usepackage[symbol]{footmisc}
\usepackage{float}

\DeclarePairedDelimiterX{\infdivx}[2]{(}{)}{%
	#1\;\delimsize\|\;#2%
}

\DeclarePairedDelimiter{\norm}{\lVert}{\rVert}

\newcolumntype{P}[1]{>{\centering\arraybackslash}p{#1}}
\newcolumntype{M}[1]{>{\centering\arraybackslash}m{#1}}

\pgfplotsset{
	compat=newest,
	xlabel near ticks,
	ylabel near ticks
}

\theoremstyle{definition}

\numberwithin{equation}{section}

\begin{document}


\title[Review of Multi-fidelity Models]{Review of multi-fidelity models}

\author[M. Giselle Fern\'andez-Godino]{}

\subjclass{ }


\thanks{This work was performed under the auspices of the U.S. Department of Energy by Lawrence Livermore National Laboratory under Contract DE-AC52-07NA27344 and supported by the LLNL-LDRD Program under Project No. 21-ERD-028.}



\maketitle

\centerline{\scshape
{M. Giselle Fern\'andez-Godino}$^{{\href{mailto:fernandez48@llnl.gov}{\textrm{\Letter}}}}$
}



\medskip

{\footnotesize
 \centerline{{Lawrence Livermore National Laboratory, Livermore, California 94550}
}}

\bigskip

 \centerline{(Communicated by Annalisa Quaini)}

\bigskip

\centerline{\emph{This paper is dedicated to Raphael T. Haftka.}}

\begin{abstract}
Multi-fidelity models provide a framework for integrating computational models of varying complexity, allowing for accurate predictions while optimizing computational resources. These models are especially beneficial when acquiring high-accuracy data is costly or computationally intensive. This review offers a comprehensive analysis of multi-fidelity models, focusing on their applications in scientific and engineering fields, particularly in optimization and uncertainty quantification. It classifies publications on multi-fidelity modeling according to several criteria, including application area, surrogate model selection, types of fidelity, combination methods and year of publication. The study investigates techniques for combining different fidelity levels, with an emphasis on multi-fidelity surrogate models. This work discusses reproducibility, open-sourcing methodologies and benchmarking procedures to promote transparency. The manuscript also includes educational toy problems to enhance understanding. Additionally, this paper outlines best practices for presenting multi-fidelity-related savings in a standardized, succinct and yet thorough manner. The review concludes by examining current trends in multi-fidelity modeling, including emerging techniques, recent advancements, and promising research directions.
\end{abstract}


\section{Introduction}
\subsection{The need for multi-fidelity modeling} \label{subsec:mot}
\begin{figure}[!ht]
	\begin{center}
		\includegraphics[width=10cm]{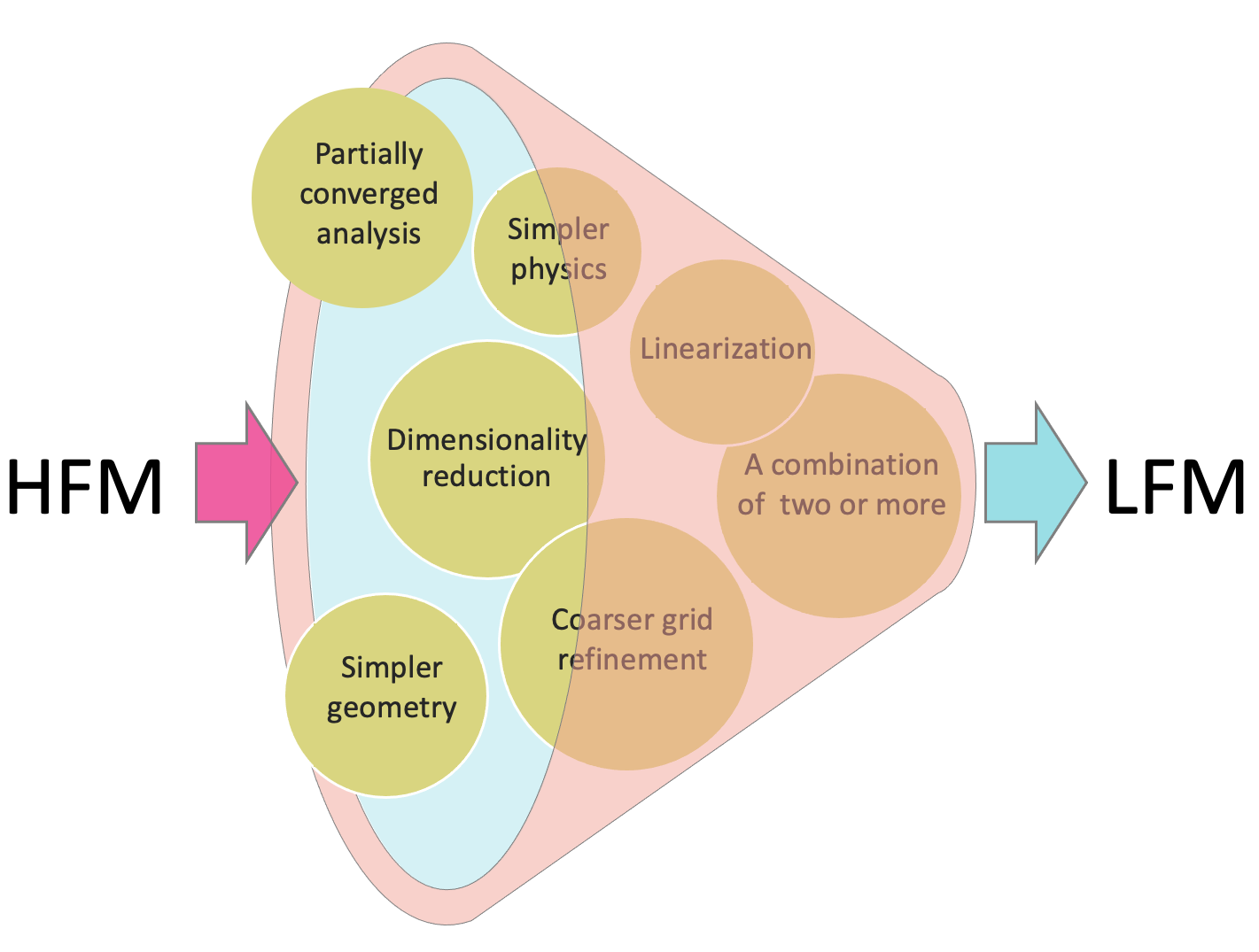}
		\caption{\label{fig:HF_LF_Liason} Connection between high-fidelity and low-fidelity models is commonly attributed to one or more of the following factors: dimensionality reduction, grid coarsening, linearization, partial convergence, reduced geometry complexity, and simplified physics.}
	\end{center}
\end{figure}

In science and engineering, high-fidelity models (HFMs) are complex systems that can make predictions with the desired accuracy. However, the cost of developing and utilizing such models is often prohibitively high, which limits their practicality for many applications. Alternatively, low-fidelity models (LFMs), their simpler and less expensive counterparts, offer a more affordable alternative. LFMs can be less accurate due to simplifications such as dimensionality reduction, linearization, use of simpler physics models, coarser domains, or partially converged results, as depicted schematically in Figure \ref{fig:HF_LF_Liason}. Note that if a model is considered to have lower or higher fidelity, it can only be determined relative to another. For instance, a three-dimensional direct numerical simulation can be considered more accurate than an analytical function evaluation but of lower fidelity than an actual field experiment. By combining different fidelities, multi-fidelity models (MFMs) aim to bridge the gap between rapid computation and high accuracy, leveraging multiple fidelities' strengths in a single model. Due to their potential to achieve a desired level of accuracy at a lower cost, since the early 2000s, MFMs have gained significant attention.

Given the variations in definitions surrounding fidelity models, specifying the conventions adopted in this survey is relevant. Here, the term \textit{multi-fidelity} refers exclusively to methods that combine at least two physical models. Consequently, articles that distinguish between a physical model and its simplified surrogate as different fidelities are not considered MFMs. Similarly, methods that utilize multi-level techniques, which replace HFMs with LFM with occasional accuracy checks, are also not classified as MFMs in this survey. These model reduction methods speed up processes such as optimization at the cost of reduced accuracy. In contrast, MFMs aim to balance accuracy and affordability.

\subsection{Survey scope and related work}
As the domain of MFMs continues to evolve, it becomes evident that comprehensive surveys are scarce. This review and that of Peherstorfer et al. (2018a) \cite{peherstorfer2018survey} stand as holistic evaluations dedicated to MFMs. While Peherstorfer et al. (2018) deeply explores the techniques for \textit{outer-loop} applications of MFMs, this review emphasizes the trade-offs between accuracy and computational cost when integrating multiple physics-based models. This work particularly focuses on surrogate-based MFMs, a concept introduced in the subsequent subsection \ref{subsec:SBMFO}. Other reviews in the literature often limit their scope to particular applications, as seen in the work of Leifsson and Koziel (2015) \cite{leifsson2015aerodynamic}, which narrows its discussions to cases of transonic airfoil design tasks. Although not a formal survey, the work of Fernández-Godino et al. (2019) \cite{fernandez2019issues} complements this review. It adopts a more critical perspective, probing into the optimal use of MFS and addressing the challenges during the process. In contrast, this review is educational and informative, presenting an expansive overview, categorization and current trends in the multi-fidelity modeling domain.

\subsection{Surrogate-based multi-fidelity modeling} \label{subsec:SBMFO}

Surrogate models (SMs) are mathematical approximations employed to mimic a system's behavior, primarily to cut down on prediction time and costs. They come in handy when faced with numerous, expensive simulations, such as those needed for optimization \cite{forrester2007multi, viana2014special, zhang2021multi} or uncertainty quantification (UQ) \cite{perdikaris2015multi, biehler2015towards, farcas2023context}.

A distinction is necessary here: A \textit{model} is based on physics-driven equations, while a \textit{surrogate model} is a data-driven mathematical representation predicting specific outcomes. To illustrate this point with an example, computational fluid dynamics or CFD codes are considered physics-based models. However, when code-generated data is used to build a mathematical function to predict a specific quantity of interest (e.g., pressure as a function of temperature, $P=f(T)$) via methods such as the least squares approach \cite{hansen2013least}, the outcome is an SM. In this review, HFMs or LFMs refer to models unless stated otherwise.

\subsection{Delving into multi-fidelity surrogate modeling}

Our primary focus is MFSM, in which architectures fuse data from different fidelity levels or their corresponding individual-fidelity SMs into a single predictor. Though the concept of surrogate modeling is assumed to be familiar to readers, a brief on the commonly used SMs in the MFSM context is provided in Appendix \ref{sec:SurrogateModels}. The selection of a suitable SM is often data-and outcome-centric. For instance, predicting a three-dimensional temporal fluid dynamics model output might favor a neural network SM because those models can handle large datasets. In contrast, a scalar density prediction could lean towards a regression or random forest model.

\subsection{Challenges and solutions in computational simulations}

Obtaining ample HFM data for an accurate SM can strain computational resources. One\break workaround is leveraging affordable LFMs or older, less computationally intensive methods. There are instances where LFMs prove cost-efficient for direct use, as highlighted by Nguyen et al. (2013) \cite{nguyen2013multidisciplinary}. Alternatively, SMs can be designed to approximate LFMs.

\subsection{Integration of various fidelity levels}
\begin{figure}[!ht]
	\begin{center}
		\includegraphics[width=12cm]{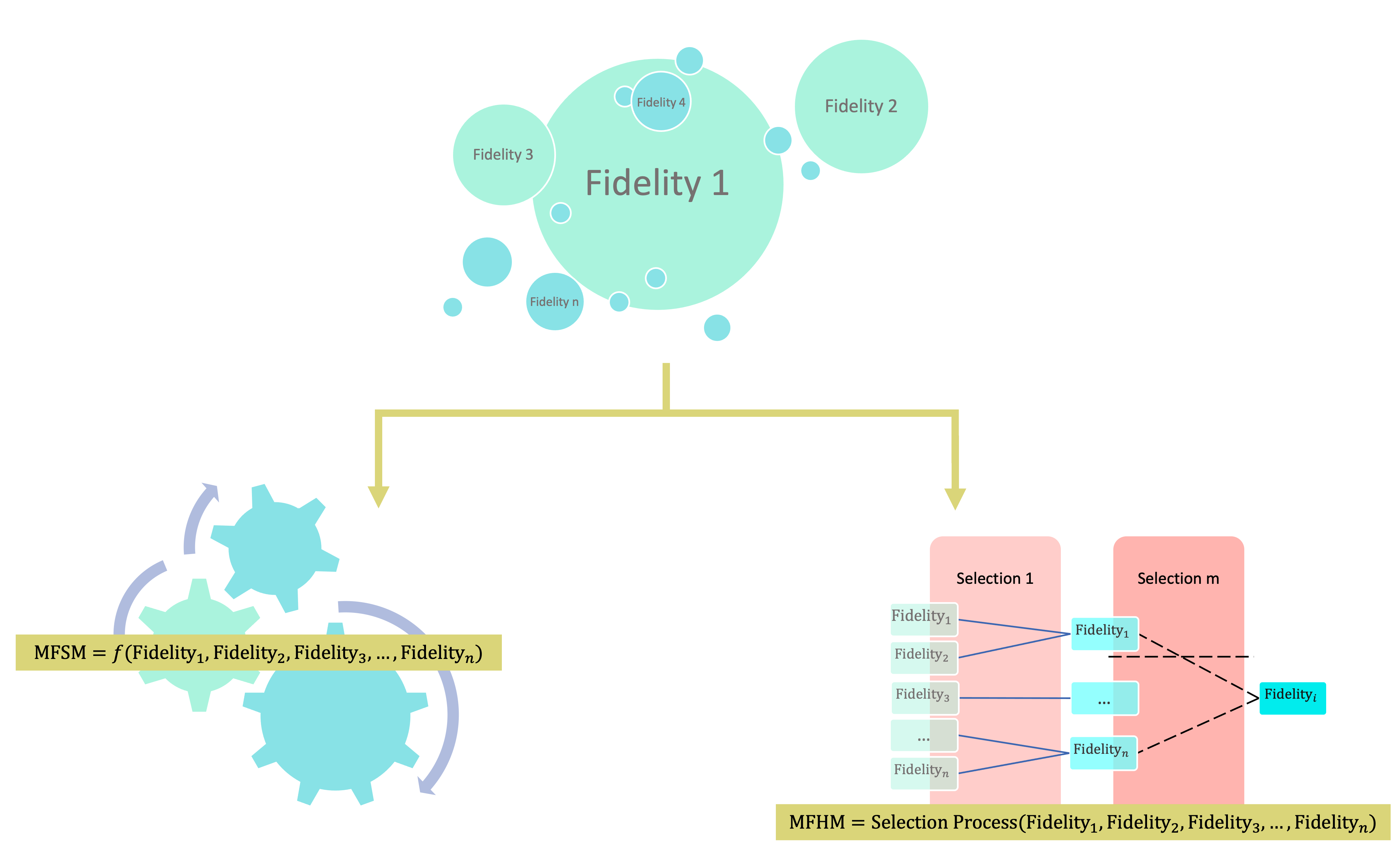}
		\caption{\label{fig:MFM} Within the frame of multi-fidelity modeling, surrogate models are commonly used to integrate information from different fidelities. When constructing a surrogate model that combines fidelities explicitly, such as co-Kriging, the resulting approach is referred to as a multi-fidelity surrogate model. In contrast, multi-fidelity hierarchical models combine fidelities without requiring to build an explicit multi-fidelity surrogate model architecture. Methods such as importance sampling fall under the multi-fidelity hierarchical category.}
	\end{center}
\end{figure}

The construction of an MFSM by integrating various fidelity levels in a single SM is not mandatory for MFMs. This alternative MFM technique is known as a multi-fidelity hierarchical model (MFHM). For instance, Choi et al. (2008) \cite{choi2008multifidelity} employed different fidelity types proficiently through adaptive sampling without constructing an MFSM. Other examples are Kalivarapu and Winer (2008) \cite{kalivarapu2008multi}, where an MFM is used for interactive modeling of advective and diffusive contaminant transport with no MFSM construction. This approach is also shown in Farcaș et al. (2023) \cite{farcas2023context}, which proposed a hierarchical context-aware multi-fidelity Monte Carlo method that optimally balances the costs of training LFMs with the costs of Monte Carlo sampling. Other examples are Giunta et al. 1995 \cite{giunta1995variable} and Zahir et al. (2013) \cite{zahir2013variable}. Figure \ref{fig:MFM} illustrates the options for constructing an MFM. In both approaches, HFMs and LFMs or their SMs are used.

\subsection{Retrospective overview of multi-fidelity model attributes}

This section offers an overview of multi-fidelity modeling attributes, drawing from the literature from the late 1980s to the late 2010s. The contributions from this period have established foundational principles that remain relevant for the scientific community \footnote{The charts are based on 157 papers from the late 1980s to the late 2010s.}.

Figure \ref{fig:Charts} provides a visual summary of selected attributes from the reviewed literature. Figure \ref{fig:Application} indicates that MFMs primarily apply in three areas: optimization (encompassing inference and inverse problems), uncertainty quantification (UQ) and optimization under uncertainty. Papers presenting general procedures without specifying applications are categorized under \textit{none}. Figure \ref{fig:Types} showcases the nature (origin) of fidelities combined, such as Physics, numerical solution accuracy, numerical models and the blending of simulations with experiments (Sim$+$exp). Further discussion on this category is available in Section \ref{sec:UnderstandingModelFidelity}.

The strategy for data assimilation in MFSM creation is distinguished by the method employed, either deterministic (DM) or non-deterministic (NDM). In brief, DMs produce fixed results for specific inputs, while NDMs can yield different results even with the same inputs because they account for uncertainties. Figure \ref{fig:Method} reveals an almost equal presence of both DMs and NDMs. This topic is expanded in Section \ref{subs:det}.

Figure \ref{fig:Year} confirms a growing interest in MFMs since their inception in the late 1980s. Figure \ref{fig:Field} throws light on the fields where MFMs gained traction, with fluid and solid mechanics emerging as dominant areas, followed by electronics, aeroelasticity and thermodynamics. Finally, Figure \ref{fig:Surr} illustrates the preferred surrogate types for MFSM development. The analysis identified basis function regression, also known as response surface models, and Kriging as the frontrunners. More details on SMa are available in Appendix \ref{sec:SurrogateModels}.

\begin{figure}[!ht]
	\centering
	\begin{subfigure}{0.45\textwidth}
		\centering
		\begin{tikzpicture}[scale=0.4]
			\pie[polar, radius=4, /tikz/every pin/.style={align=center},style={lines}, /tikz/nodes={text=black!80}, text=pin, explode=0.1, rotate=270,color={teal!10,teal!20,teal!30,teal!40}]{56/Optimization, 23/Optimization\\ under \\ Uncertainty, 8/None, 13/Uncertainty \\Quantification}
		\end{tikzpicture}
		\caption{Application.}\label{fig:Application}
	\end{subfigure}
	\hspace{1cm}
	\begin{subfigure}{0.45\textwidth}	
		\begin{tikzpicture}[scale=0.4]
			\pie[polar, radius=4, /tikz/every pin/.style={align=center},style={lines}, /tikz/nodes={text=black!80}, text=pin,explode=0.0,rotate=150, color={teal!10,teal!20,teal!30,teal!40,teal!50}]{5/Sim+exp, 13/Numerical\\models, 15/Others, 27/Numerical\\solution\\accuracy, 40/Physics}
		\end{tikzpicture}
		\caption{Types of fidelity.}\label{fig:Types}
	\end{subfigure}
	\vskip\baselineskip
	\begin{subfigure}{0.45\textwidth}
		\begin{tikzpicture}[scale=0.4]
			\pie[polar, radius=3, /tikz/every pin/.style={align=center},style={lines}, text=pin,explode=0.0,color={teal!10,teal!20,teal!30}]{30/Non-\\deterministic\\methods, 33/None, 37/Deterministic\\methods}
		\end{tikzpicture}
		\caption{\small Method.}\label{fig:Method}
	\end{subfigure}
	\hfill
	\begin{subfigure}{0.46\textwidth}
		\begin{tikzpicture}[scale=0.4]
			\pie[polar, radius=4, /tikz/every pin/.style={align=center}, style={lines}, /tikz/nodes={text=black!80}, text=pin,explode=0.0,rotate=-20,color={teal!10,teal!20,teal!30,teal!40,teal!50}]{2/{1987-1992}, 12/{1993-1998}, 13/{1999-2004}, 31/{2005-2010}, 42/{2011-2016}}
		\end{tikzpicture}
		\caption[]{{\small Year published.}}\label{fig:Year}
	\end{subfigure}
	\vskip\baselineskip
	\begin{subfigure}{0.45\textwidth}
		\begin{tikzpicture}[scale=0.4]
			\pie[polar, radius=4, /tikz/every pin/.style={align=center},style={lines}, /tikz/nodes={text=black!80},text=pin,explode=0.0,rotate=20,color={teal!10,teal!20,teal!30,teal!40}]{46/Fluid\\Mechanics, 25/Solid\\mechanics, 13/Other, 16/None}
		\end{tikzpicture}
		\caption{\small Field.}\label{fig:Field}
	\end{subfigure}
	\hfill
	\begin{subfigure}{0.45\textwidth}
		\begin{tikzpicture}[scale=0.4]
			\pie[polar, radius=3, /tikz/every pin/.style={align=center},style={lines}, /tikz/nodes={text=black!80},text=pin,explode=0.0,rotate=30,color={teal!10,teal!20,teal!30,teal!40}]{31/Basis function\\regression, 25/Kriging, 36/None, 8/Others}
		\end{tikzpicture}
		\caption{\small Surrogate model.}\label{fig:Surr}
	\end{subfigure}
	\caption{Proportions of different attributes found in the reviewed multi-fidelity literature \cite{fernandez2019issues}.}
	\label{fig:Charts}
\end{figure}

\subsection{Review structure}

The organization of this review is as follows. Section \ref{sec:UnderstandingModelFidelity} offers an in-depth discussion of what constitutes model fidelity. It explores the correlation between various fidelities, delves into their hierarchical structure and categorization and briefly discusses low-fidelity model classifications. Section \ref{sec:DomainsOfApplication} articulates the wide range of applications where MFMs are employed, reflecting their versatility and broad applicability. Then, Section \ref{sec:CombiningFidelities} delves into methodologies of fusing or hierarchically leveraging multiple fidelities. This section also discusses and compares deterministic and stochastic methods.

Section \ref{sec:Reporting} underscores the relevance of good practice when reporting cost savings in MFMs, advocating for open-source dissemination and standardized benchmarking. In Section \ref{sec:CurrentTrends}, contemporary advancements in the field are outlined, focussing on current challenges such as model interpretability and incorporating physics into MFMs. The paper then transitions into Section \ref{sec:FutureDirections}, which extrapolates current trends to forecast emergent research avenues and methodologies. Section \ref{sec:Conclusion} summarizes the core findings and contributions of the paper.

The appendices augment the main text, starting with Appendix \ref{sec:Nomenclature}, a glossary of terms and concepts fundamental to the multi-fidelity domain. Subsequently, Appendix \ref{sec:DesignOfExperiments} discusses various sampling techniques suited for MFMs, spanning traditional to adaptive strategies. Appendix \ref{sec:SurrogateModels} is an introductory description of several highly used surrogate models, particularly in the MFM domain. Finally, Appendix \ref{sec:ToyProblems} features a collection of illustrative MFM toy problems designed to translate theoretical discussions into practical insights.

\section{Understanding model fidelity} \label{sec:UnderstandingModelFidelity}

\begin{figure}[ht]
	\begin{center}
		\includegraphics[width=12cm]{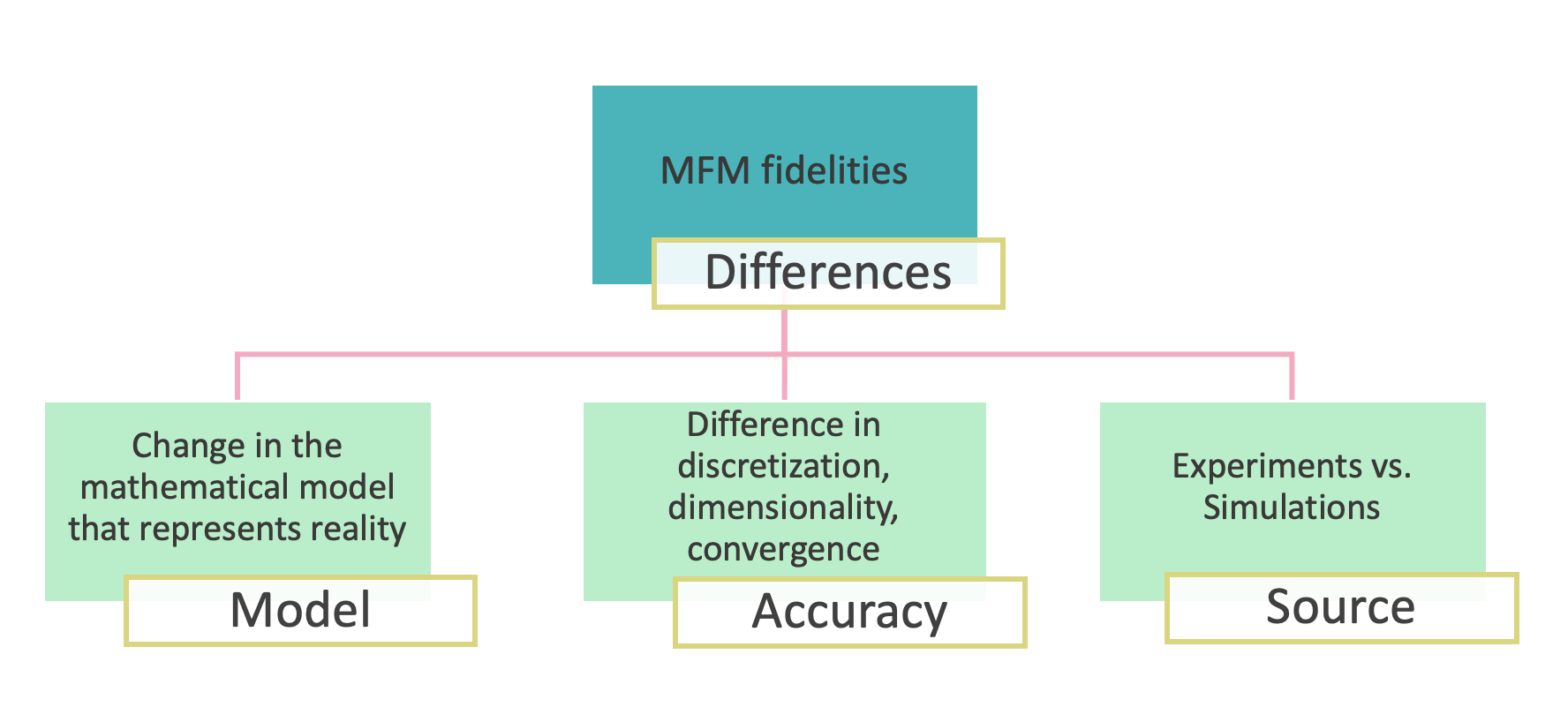}
		\caption{\label{fig:LFvsHF} Main differences between fidelities found in the literature.
		}
	\end{center}
\end{figure}

The current study reviews the literature and identifies different types of fidelities commonly associated with three principal categories: \textit{model}, \textit{accuracy} and \textit{source} as depicted in Figure \ref{fig:LFvsHF}. The \textit{Model} category simplifies the mathematical representation of the physical phenomenon, typically by simplifying the differential equations being solved or the numerical model or reducing dimensions. \textit{Accuracy} refers to changes in the discretization of the model, such as using smaller grid elements or shorter time steps for HFMs. \textit{Source} is associated with incorporating experimental results in addition to simulations, which are regarded as having the highest level of fidelity.

\subsection{Relationship between fidelities}

Understanding the relationship between different levels of fidelity is critical to achieving good MFM performance (Forrester et al., 2007 \cite{forrester2007multi}). As expected, a simple relationship between the LFM and the HFM is highly associated with how well an MFM can (or has the potential to) approximate the behavior of the HFM. For instance, the difference between LF and HF outputs may follow a consistent trend across the design space, suggesting a linear discrepancy (Kennedy and O'Hagan, 2000 \cite{kennedy2000predicting}). A simple discrepancy function is desirable in SM techniques, especially for models such as co-Kriging, where it is explicitly considered, enabling the blending of data across fidelity levels (Alexandrov et al., 2001 \cite{alexandrov2001approximation}). Some literature even suggests using first a multiplicative correction factor that makes the discrepancy \textit{more suitable to be fitted} by reducing its \textit{bumpiness}, which highly depends on \textit{waviness} and \textit{variation} (Park et al., 2017 \cite{park2017remarks} and 2018 \cite{park2018low}).

Quantifying a \textit{more suitable to be fitted} relationship between fidelities is not straightforward. It may also involve cross-validation or correlation considerations, metrics that can provide insights into the reliability of the LFM as a predictor for the HFM (Le Gratiet, 2013 \cite{le2013multi}). A strong positive correlation often aids in effective information transfer between models, strengthening computational efficiency without sacrificing accuracy (Qian et al., 2008 \cite{qian2008gaussian}). Moreover, understanding this correlation in probabilistic analysis assists in precise UQ, affecting the predictions' uncertainty bounds (Peherstorfer et al., 2018b) \cite{peherstorfer2018multifidelity}). Modern methods now employ adaptive strategies adjusting model weights based on observed correlations, maximizing efficiency (Chaudhuri and Willcox, 2016 \cite{chaudhuri2016multifidelity}).

\subsection{Number and hierarchy of fidelities}

The literature typically demonstrates the construction of MFMs with two fidelities. However, most MFMs are not constrained in such a way. Many MFMs can harness (or can be extended to) more than two fidelities, as evidenced by the works of Huang et al. (2006) \cite{huang2006sequential}, Forrester et al. (2007) \cite{forrester2007multi}, Qian et al. (2008) \cite{qian2008bayesian}, and Goh et al. (2013) \cite{goh2013prediction},  Kandasamy et al. (2016) \cite{kandasamy2016gaussian} and others.

Establishing the superiority of one fidelity over another can be a complex endeavor. In some scenarios, the choice is relatively straightforward. For instance, when two simulations use an identical model but differ in grid resolution, it is generally accepted that the one with the refined grid offers superior fidelity due to its enhanced ability to capture finer details (Roache, 1998 \cite{roache1998verification}). However, other scenarios present more complex comparisons. For example, when considering a one-dimensional model with a refined grid against a three-dimensional model with a coarser grid, the decision is not merely based on the grid's resolution. The dimensionality of the model itself also plays a significant role. Therefore, practitioners should acknowledge that the superiority of a fidelity level is context-dependent and can vary based on the simulation's objectives and constraints (Peherstorfer et al., 2018a \cite{peherstorfer2018survey}).

\subsection{Classification of low fidelity models}

As Figure \ref{fig:HF_LF_Liason} schematically shows, there are multiple ways to obtain a LFM. The field of computational modeling continually evolves, often leading to diverse methodologies that aim to achieve similar objectives. One such classification, detailed by Peherstorfer et al. (2018a) \cite{peherstorfer2018survey}, categorizes LFMs into three categories. First, the \textit{Simplified Models} reduce the complexity found in HFMs, providing computational efficiency, especially in scenarios where a general understanding is prioritized over details. Second, \textit{Data-fit Models}, rather than representing a reduced complexity, are primarily grounded in SM techniques. These models fit data from HFMs or actual observations to deliver expedited outcomes. As mentioned, this survey does not consider Data-fit Models as a fidelity. Lastly, \textit{Projection-based Models} project high-dimensional data onto a reduced-dimensional space. These models capture the critical dynamics of the studied phenomena without the extensive computational demands of full simulations. More details on this category are expanded in Appendix \ref{sec:SurrogateModels}. Each classification showcases the versatility of LFMs in addressing a wide range of challenges in computational science.

\section{Domains of application} \label{sec:DomainsOfApplication}

The survey's exploration into the applications of MFMs predominantly identified two domains: fluid mechanics and solid mechanics. A spectrum of fidelity types in fluid mechanics emerges, from analytical expressions and empirical relations to more complex categories like numerical linear approximations and Euler approximations. Furthermore, the distinction between simulations and experiments and variations such as coarse versus refined analyses are key highlights. Specificities of these fidelities and their applications are detailed in Appendix \ref{sec:tables} under Table \ref{tab:types_Fluids}. Additional fidelity categories within fluid mechanics, including dimensionality, analysis refinement level and transient versus steady states, can be found in Table \ref{tab:extra_Fluids} within  Appendix \ref{sec:tables}.

Solid mechanics, on the other hand, centers on fidelities associated with mesh density, material models and temperature. A more exhaustive list, including unique fidelity classifications such as the ones showcased in Kim et al. (2007), is available in Table \ref{tab:types_Solids} of  Appendix  \ref{sec:tables}. Additional categories specific to solid mechanics, like dimensionality or boundary condition simplifications, are further elaborated in Table \ref{tab:extra_Solids}.

Further exploration of other domains, from electronics to robotics, reveals diverse fidelity determination methods. For instance, electronics often gravitate toward coarse vs. refined analyses, while in robotics, the complexity of robot resources becomes the deciding factor. Analytical function studies and methods focused on uncertainty analysis, as covered by researchers like Robinson et al. (2006) and Burton and Hajela (2003), offer yet another layer of depth to this exploration. Readers are directed to the comprehensive collection in Appendix \ref{sec:tables} for more details on these categories and their associated studies.

\section{Combining fidelities} \label{sec:CombiningFidelities}

This section discusses different methods for combining fidelities. The objective is to capitalize on the virtues of each fidelity level, aiming for a model that is both accurate and computationally efficient. Our discussion will concentrate on two predominant approaches: multi-fidelity surrogate models (MFSMs) and multi-fidelity hierarchical models (MFHMs). In MFSMs, algebraic surrogates boost LFMs performance by leveraging data from HFM predictions. Conversely, MFHMs opt for fidelity levels based on specific criteria, as illustrated in Figure \ref{fig:MFM}. This section further examines the four correction paradigms employed in MFSMs: multiplicative correction, additive correction, comprehensive correction, and space mapping. Real-world applications and illustrative examples of these techniques will also be provided to offer readers more profound insights into their practical utility.

\subsection{Fusion vs. hierarchy} \label{sub:vs}

As mentioned, the comprehensive literature review identifies two dominant strategies in MFM: MFSMs and MFHMs. As shown in Figure \ref{fig:Mult_w_o_surr}, MFSMs historically dominated the landscape, comprising most research articles, whereas MFHMs accounted for a substantial minority. MFHMs set themselves apart by employing fidelities selectively, guided by specific optimization criteria. This approach contrasts the more encompassing architecture of MFSMs, which aim for a broad integration of low- and high-fidelity data. Although the reviewed papers considered in Figure \ref{fig:Mult_w_o_surr} only cover literature up to 2016, these findings remain relevant in shaping current trends. In fact, recent observations point to a growing inclination towards MFHMs. This shift is enabled by advancements in computational capabilities and optimization algorithms, permitting more sophisticated approaches that were previously computationally prohibitive.

\begin{figure}[!ht]
	\centering
	\begin{tikzpicture}[scale=0.8]
		\pie[polar, /tikz/every pin/.style={align=center}, /tikz/nodes={text=black!80}, style={lines}, text=pin,explode=0.1,rotate=0,color={teal!10,teal!30}]{33/MFHM,67/MFSM}
	\end{tikzpicture}
	\caption{Prevalence of multi-fidelity surrogate models over multi-fidelity hierarchical models in surveyed literature up to 2016. Recent trends suggest a growing shift towards hierarchical models due to advancements in computing and algorithms.}
	\label{fig:Mult_w_o_surr}
\end{figure}

Several studies illustrate the diversity of applications and methods within\break MFHMs. Christen and Fox employed LFMs in initial Markov Chain Monte Carlo sampling and shifted to HFMs when specific criteria were met \cite{christen2005markov}. Similarly, Rethore et al. (2005) started with cost-efficient models and escalated in complexity and resolution only as required \cite{rethore2011topfarm}. In contrast, Narayan et al. (2014) used a stochastic collocation method to identify which data points should be high-fidelity, guided by low-fidelity insights \cite{narayan2014stochastic}. Peherstorfer et al. (2016) utilized an importance sampling method rooted in LFMs to determine optimal high-fidelity sampling points \cite{peherstorfer2016multifidelity}. These applications and those by Burton and Hajela (2003) \cite{burton2003variable}, Choi et al. (2005) \cite{choi2005multi} and Singh and Grandhi (2010) \cite{singh2010mixed} showcase the adaptability of MFHMs. However, as MFHMs become increasingly complex, they risk evolving into \textit{black boxes}, where the intricacies of data integration and decision-making are obscured. This lack of transparency could compromise the interpretability of the model, which is especially relevant for applications requiring rigorous validation and verification.

The survey by Peherstorfer et al. \cite{peherstorfer2018survey} provides an alternative classification for techniques used to integrate multi-fidelity data, namely, \textit{adaptation}, \textit{fusion} and \textit{filtering}. MFHM belongs to the filtering category, while MFSM approaches fit into fusion and adaptation. The adaptation technique is a bridge between low- and high-fidelity data. Its primary goal is to fine-tune the low-fidelity data, ensuring it aligns more closely with high-fidelity information. Some standard techniques used are calibration, bias correction, space mapping and data transformation \cite{rayas2016power, bandler2004space}. Fusion, on the other hand, harnesses the unique strengths inherent in each level of fidelity by merging information from multiple fidelity levels in a single SM, leveraging tools from statistical analyses \cite{forrester2008engineering} to advance machine learning methodologies \cite{pawar2022multi}. Lastly, filtering prioritizes the most pertinent information from various fidelity levels. In doing so, it effectively downplays or excludes data found to be less significant or informative \cite{narayan2014stochastic, peherstorfer2016multifidelity}.

\subsection{Multi-fidelity surrogate models} \label{subsub:avail}

This survey focuses mainly on MFSMs, which aim to integrate multiple fidelities into a unified SM. MFSMs enhance the accuracy of LFMs by incorporating data from HFMs. They primarily employ four correction paradigms: multiplicative correction, additive correction, comprehensive correction and space mapping. In scenarios where the LFM and HFM have different parameters or dimensions, problem-specific transformations may be necessary, as discussed in works by Robinson et al. (2008) \cite{robinson2008surrogate} and Koziel et al. (2009) \cite{koziel2009accelerated}.

\subsubsection{Additive and multiplicative corrections} \label{subsub:addmult}

Two prominent methods to enhance the accuracy of LFMs involve additive and multiplicative corrections. The additive approach adds an adjustment term to the LFM prediction to align it more closely with the desired HFM output. This term is calculated by analyzing the discrepancies between low-fidelity and high-fidelity datasets. In the multiplicative approach, a correction factor, based on the ratio between low-fidelity and high-fidelity data points, is multiplied by the LFM output to better match the HFM predictions.

MFSMs employing additive correction can be succinctly expressed using the following equation:
\begin{equation}\label{sum}
	\hat{y}_{\text{HF}} = y_{\text{LF}}(\mathbf{x}) + \delta(\mathbf{x}),
\end{equation}

\noindent where, \( y_{\text{LF}}(\mathbf{x}) \) represents the LFM, which an LFSM can replace if the computational expense is a concern. Meanwhile, \( \delta(\mathbf{x}) \) is an additive correction or discrepancy model, capturing the difference between the HFM and the LFM.

In the case of a MFSM that employs multiplicative correction, the estimated HFM can be mathematically represented as follows:
\begin{equation}\label{division}
	\hat{y}_{\text{HF}} = \rho(\textbf{x}) \cdot y_{\text{LF}}(\textbf{x}),
\end{equation}

\noindent where \(\rho(\textbf{x})\) serves as the multiplicative correction factor, essentially an SM created from the ratio between the HFM and the LFM. Again, if the LFM proves too computationally expensive, it can be substituted with an LFSM as a more efficient alternative.

MFSMs are frequently employed in aerodynamic optimization to save computational time. Various correction methods have been explored to build these models. For example, Alexandrov et al. (2001) \cite{alexandrov2001approximation} used multiplicative corrections in aerodynamic optimization, while Balabanov et al. (1998) \cite{balabanov1998multifidelity} assessed both additive and multiplicative approaches for similar tasks. Forrester et al. (2006) \cite{forrester2006optimization} applied an additive correction to improve partially converged results based on fully converged data. For further literature on additive and multiplicative corrections in MFSMs, please refer to Tables \ref{tab:DF} and \ref{tab:NDF} in Appendix \ref{sec:tables}. Hands-on examples illustrating these concepts are provided in Appendix \ref{sec:ToyProblems}, and the open-source code to replicate it is available in the supplementary material.

\subsubsection{Comprehensive corrections} \label{susub:compre}

Comprehensive correction incorporates both additive and multiplicative corrections in the same MFSM. It aims to simultaneously address discrepancies in both the magnitude and trend of the LFM compared to the HFM. Comprehensive correction techniques are more challenging but can be better predictors in most cases. These approaches often involve statistical analysis, regression models or data-driven approaches to estimate the appropriate adjustments for the low-fidelity data. One simple yet widely used comprehensive correction is defined as follows:
\begin{equation}
	\hat{y}_{HF}=\rho(\textbf{x}) \cdot y_{LF} (\textbf{x}) + \delta(\textbf{x})
\end{equation}

\noindent where $\rho(\textbf{x})$ represents the multiplicative correction surrogate, and $\delta(\textbf{x})$ represents the additive correction surrogate. A common approach is to set the multiplicative factor $\rho$ as a constant and to use an SM to approximate the additive correction, as has been demonstrated in previous studies such as Keane (2012)\cite{keane2012cokriging}, Perdikaris et al. (2015)\cite{perdikaris2015multi} and Zhang et al. (2018) \cite{zhang2018multifidelity}. However, Qian et al. (2008) \cite{qian2008bayesian} proposed a comprehensive correction method with a non-constant $\rho(x)$. Gano et al. in 2005 \cite{gano2005hybrid} devised an alternative \text{hybrid} comprehensive correction technique where the high-fidelity prediction can be defined as follows:
\begin{equation}
	\hat{y}_{HF}=w(\textbf{x}) \cdot \rho(\textbf{x}) \cdot y_{LF} (\textbf{x}) + (1-w(\textbf{x})) [y_{LF}(\textbf{x}) + \delta(\textbf{x})],
\end{equation}

\noindent where $w(\textbf{x})$ is a weight function. A weight function assigns greater importance to specific data points while training an SM. This technique has been adopted in numerous research studies, including those conducted by Zheng et al. (2013) \cite{zheng2013hybrid} and Fischer et al. (2017) \cite{fischer2017bayesian}. In Tables \ref{tab:DF} and \ref{tab:NDF} presented in Appendix \ref{sec:tables}, the interested reader can locate additional references that illustrate the construction of MFSMs utilizing comprehensive corrections. The reader can also find a practical example illustrating these concepts in Appendix \ref{sec:ToyProblems}. The repository containing the open-source code to reproduce this toy example is included in the supplementary material.

\subsubsection{Space mapping} \label{susub:space}

Space mapping is an adaptation approach that iteratively aligns the LFM to the HFM in the parameter space. The primary motivation for this method is that if the LFM can closely resemble the HFM in a particular design space, optimization can be performed much more efficiently using the LFM while still maintaining the accuracy of the HFM.
The mapping is typically derived through data fitting or model calibration techniques. Bandler et al. \cite{bandler1994space, bandler2013have} introduced space mapping in the early 1990s. The fundamental concept behind this technique is to generate an appropriate transformation of the vector of HF input parameters, $\textbf{x}_{HF}$, to the vector of LF input parameters, $\textbf{x}_{LF}$, such that
\begin{equation} \label{eq:SMap}
	\textbf{x}_{HF} \approx F(\textbf{x}_{LF}) = \hat{\textbf{x}}_{HF}.
\end{equation}

The purpose of this iterative process is to allow the vectors $\textbf{x}_{HF}$ and $\textbf{x}_{LF}$ to have varying dimensions. Although it is not necessary, it is desirable for $\textbf{F}$ to be invertible. The objective is to ensure that the response of the HFM, $\textbf{y}_{HF}(\textbf{x}_{HF})$, and the response of the LFM, $\textbf{y}_{LF}(\textbf{x}_{LF})$, fulfill the following condition:
\begin{equation}
	\norm{y_{HF} (\textbf{x}_{HF}) - y_{LF} (\hat{\textbf{x}}_{HF})}\leq \epsilon
\end{equation}

\noindent within some local region, where $\norm{ \cdot }$ is a suitable norm, and $\epsilon$ is a tolerance setting. Space mapping has only been found in publications using DMs.

The first review paper on the space mapping method was published ten years after its implementation \cite{bandler2004space}, and a second survey was published two decades later \cite{rayas2016power}. The concept of space mapping has been expanded to include other techniques, such as aggressive space mapping \cite{bandler1995electromagnetic}, trust regions \cite{bakr1998trust}, artificial neural networks (ANNs) \cite{bakr2000neural}, implicit space mapping \cite{bandler2004implicit}, neural-based space mapping \cite{zhang2005efficient, zhang2008statistical}, inverse problems \cite{rayas2005linear}, corrected space mapping \cite{robinson2008surrogate} and tuning space mapping \cite{koziel2009accelerated}. Further literature regarding the utilization of space mapping to construct MFSMs can be found in Table \ref{tab:DF} of Appendix \ref{sec:tables}.

\subsection{Deterministic vs. stochastic} \label{subs:det}

\subsubsection{Non-deterministic methods}

Non-deterministic methods (NDMs), or stochastic methods, are SMs where variability and uncertainty are explicitly considered. In practice, this variability means that the output is not associated with a single value but follows some probability distribution for a given set of input variables. Instead of fixing the basis functions and their coefficients, NDMs introduce a layer of uncertainty either in the functional form or the coefficients themselves. Due to their complex implementation, NDMs are particularly well-suited for systems characterized by inherent variability or where uncertainty quantification is critical to decision-making. For instance, such methods are often employed in scenarios where inaccurate modeling could result in substantial financial losses or compromise human safety. Gaussian process models that employ a non-zero \textit{nugget effect} fall under NDMs \cite{ankenman2010stochastic, williams2006gaussian}. Methods using Monte Carlo simulations to recreate variability are also considered NDMs \cite{metropolis1949equation, robert2004monte}. Multiple studies suggest that compared to their deterministic counterparts, NDMs offer higher accuracy levels in predictions \cite{keane2012cokriging, park2017remarks}.

Gaussian processes are non-parametric statistical frameworks, providing a function distribution capable of making generalizable predictions on previously unobserved data \cite{kennedy2000predicting, williams2006gaussian}. Specifically, a Gaussian process is a collection of random variables so that any joint distribution of these variables follows a Gaussian distribution. In this work, Kriging is considered a specific form of Gaussian process regression, where the primary objective is to establish an interpolating function through a specific kernel and set of optimization procedures \cite{le2013multi, le2014bayesian}. While the terms Gaussian process and Kriging are often used interchangeably in the literature, it is worth noting that there are scenarios, such as using custom likelihood functions or multi-output problems, where the broader flexibility of Gaussian Processes is generally not encompassed by traditional Kriging methods \cite{le2014bayesian}. Furthermore, while Kriging models each fidelity level separately, co-Kriging, the multi-fidelity extension of Kriging, can exploit the relationships between different fidelity levels through data integration in an MFSM to provide a more accurate global SM. This data fusion leads to more efficient low- and high-fidelity data utilization in constructing a unified predictive model \cite{le2015cokriging, le2013multi}.

Traditional co-Kriging models often assume that the uncertainties across different fidelities are independent, or at least conditionally independent, given the inputs. This assumption makes the model easier to interpret and manage computationally. However, in applications with reason to believe that the errors at different fidelities are correlated, researchers are exploring ways to account for possible correlations between uncertainties at different fidelity levels. For instance, one can introduce cross-covariance terms into the co-Kriging equations to capture the dependency between different fidelities. Accounting for these correlations can improve the model's predictive performance, making the model more complex and computationally intensive \cite{qian2008bayesian, le2014recursive, forrester2009recent}. Co-Kriging applications can be found in Kennedy and O'Hagan (2000) \cite{kennedy2000predicting}, LeGratiet (2012) \cite{le2015cokriging}, LeGratiet (2013) \cite{le2013multi} and LeGratiet (2014) \cite{le2014bayesian}. Additional examples of NDMs, as well as the methodologies used to combine fidelities, namely, additive correction, multiplicative correction, comprehensive correction and space mapping, are detailed in  Appendix \ref{sec:tables} in Table \ref{tab:NDF}.

For outer-loop applications, specifically UQ, NDMs need a method of statistical inference in order to treat parameter uncertainties, therefore avoiding the costly standard Monte Carlo sampling \cite{robert1999monte,hammersley2013monte}. A prevalent NDM is the Bayesian framework, which employs Bayes' theorem to derive the posterior distribution for the model's unknown parameters. This distribution is conditioned on both the prior distribution for the parameters and the likelihood of the observed data. Gaussian processes have become popular and widely used classes of NDMs due to their flexibility and convenience in incorporating prior knowledge about the data \cite{kennedy2001bayesian}. Another great advantage of Gaussian processes is the simplicity with which new data can be added without degrading the performance of the already trained areas. Alternative methods to the Gaussian processes have also been proposed, such as the approach presented by Koutsourelakis (2009) \cite{koutsourelakis2009accurate}, where non-Gaussian distributions are used to model the uncertainties, desirable when the variability is not well represented by a normal distribution. However, while Gaussian process-based models have succeeded in low-dimensional contexts, they are unsuitable for high-dimensional problems or large datasets.

Calibration is a widely used technique in engineering for fine-tuning the parameters of simulations or physical models to align their outputs with observed real-world data better. Extensive research in the engineering domain, such as works by Kosonen and Shemeikka (1997) \cite{kosonen1997use}, Owen et al. (1998) \cite{owen1998parametric}, Lee et al. (2008) \cite{lee2008computer}, McFarland et al. (2008) \cite{mcfarland2008calibration}, Coppe et al. (2012) \cite{coppe2012using} and Yoo et al. (2013) \cite{yoo2013probabilistic}, has been dedicated to leveraging calibration to enhance simulation predictions. Bayesian calibration takes this further by employing a probabilistic framework for parameter estimation \cite{higdon2004combining, kennedy2001bayesian}. Calibration faces scrutiny as tuning model parameters for a specific scenario may inadvertently compromise its general predictive capabilities across broader contexts.

In a standard Bayesian calibration framework, the primary focus is on using a probabilistic model to estimate uncertain parameters in a simulation model so that the model output aligns more closely with observed data. Given the observed data, the primary goal is to find the posterior distribution of the uncertain model parameters. Bayesian calibration generally focuses on single-fidelity models and does not explicitly account for model discrepancy. Kennedy and O'Hagan (2001) \cite{kennedy2001bayesian} extended Bayesian calibration by introducing the notion of \textit{model discrepancy} and considering calibration parameters as non-physical hyper-parameters. In other words, Kennedy and O'Hagan's approach can be framed under \textit{Bayesian calibration along with comprehensive correction}, allowing for a more refined approach to managing the biases and uncertainties inherent to various fidelity levels. Subsequent research, including works by Qian et al. (2008) \cite{qian2008bayesian} and Biehler et al. (2015) \cite{biehler2015towards}, has explored the diverse applications of this methodology.

Polynomial chaos is another widely used method for UQ, sensitivity analysis, and calibration. Polynomial chaos is a mathematical framework representing and propagating uncertainties through computational models. Polynomial chaos expands the uncertain model inputs in orthogonal polynomials and projects the model output onto the same polynomial basis. This transformation efficiently estimates output uncertainties, sensitivity indices, and other statistical properties. In the MFM context, polynomial chaos is an SM technique that offers a probabilistic way to combine model outputs from different fidelities \cite{ng2012multifidelity, padron2014multi}.

\subsubsection{Deterministic methods}

Deterministic methods (DMs) are SMs with a single, fixed relationship between the input and output variables. In other words, the output is predetermined for a given set of input variables and will not change if the model is rerun with the same inputs. Note that DMs are not necessarily injective (one-to-one) operators because they allow multiple inputs to be mapped to the same outputs. These models are often used when the underlying process being modeled is considered well-understood and not influenced by random variables. DMs are characterized by their reliance on a predefined set of basis functions. These methods aim to determine the coefficients of these basis functions by employing optimization techniques that minimize the discrepancy between the observed data and the predictive model \cite{vitali2002multi, goel2009comparing}. DMs offer the advantage of not requiring an uncertainty structure, making them broadly applicable across different surrogates. Standard DMs include polynomial regression \cite{draper2014applied}, Gaussian processes when using a nugget effect of zero \cite{sacks1989design} and various types of neural networks \cite{goodfellow2016deep}. Further examples of DMs, along with the strategies for fidelity combination (additive correction, multiplicative correction, comprehensive correction and space mapping), are found in Table \ref{tab:DF} of  Appendix \ref{sec:tables}.

\subsubsection{Analysis and comparison}

\begin{figure}[!ht]
\vspace*{10pt}	\begin{center}
		\includegraphics[width=12.5cm]{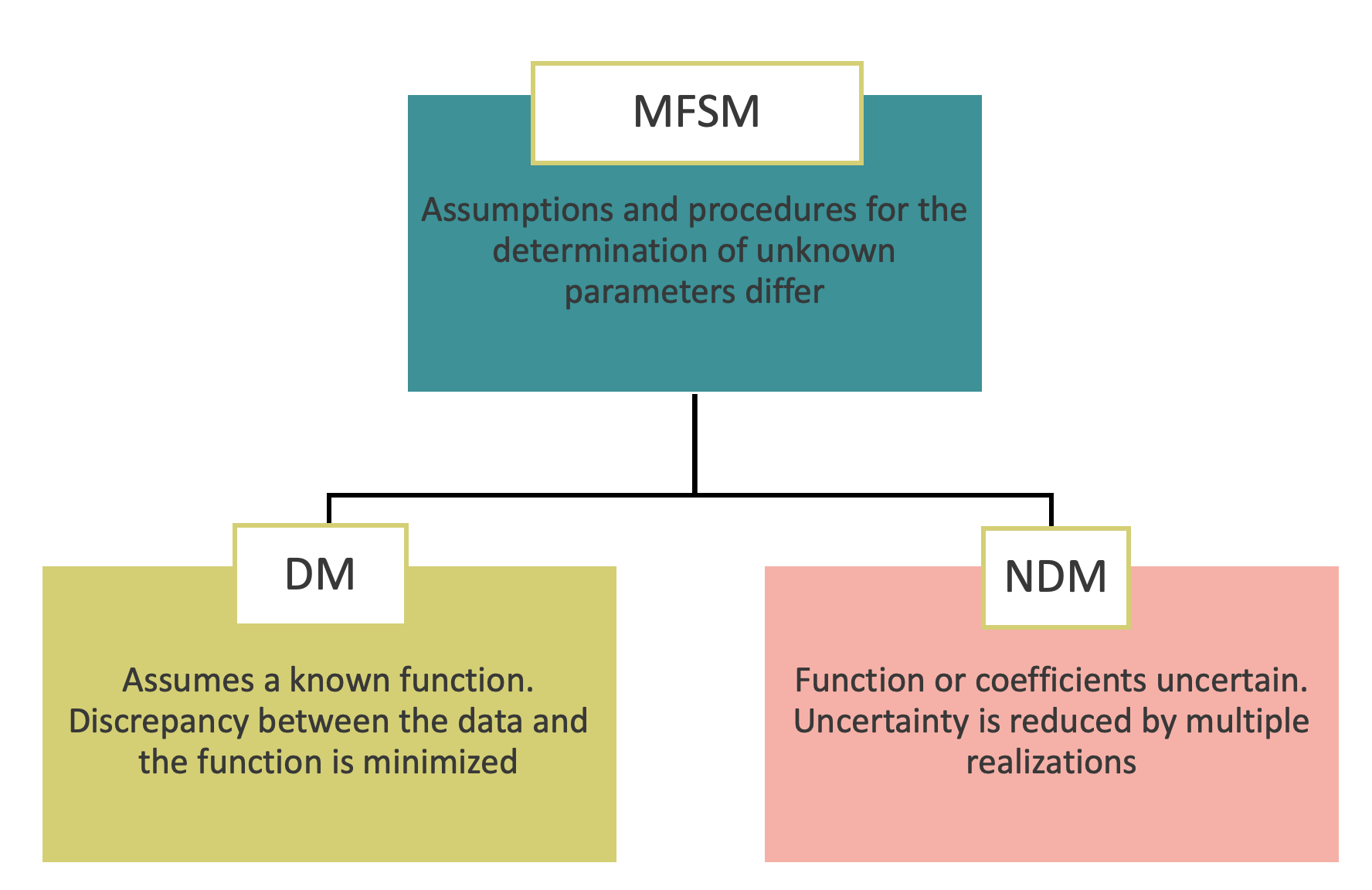}
		\caption{\label{fig:MFSM} Multi-fidelity surrogate models' parameters are inferred utilizing either deterministic or non-deterministic methodologies contingent upon the underlying presumptions of the unknown parameters.}
	\end{center}
\end{figure}

The contrasting approaches of DMs and NDMs are visually summarized in Figure \ref{fig:MFSM}, highlighting the fundamental assumptions about the unknown parameters in both methods. This classification serves as a foundation for understanding the advantages and limitations of each approach and provides a framework for selecting the most appropriate modeling strategy for a given application.

In the context of multi-fidelity, DM parameters are estimated by minimizing the difference between the predictions of the LFSM ($\hat{y}_{LF}$) and the HFSM ($\hat{y}_{HF}$) at the high-fidelity data points. On the other hand, NDMs, such as co-Kriging, estimate parameters that simplify the discrepancy function $\delta$ as much as possible, even if this increases its magnitude. By simplifying $\delta$, the accuracy of the discrepancy surrogate can be improved beyond that achieved by simply minimizing the discrepancy \cite{park2018low}.

Figure \ref{fig:combination} shows that DMs are employed in constructing MFSMs 55\% of the time, compared to 45\% that utilize NDMs \footnote {The results are based on a thorough analysis performed considering literature up to the year 2016. Nevertheless, the analysis remains pertinent for understanding foundational methods and historical trends in MFSM development}. The figure also provides insights into the prevalence of specific methods in MFSM construction, as detailed in Section \ref{subsub:avail}. The data indicates that multiplicative methods are favored mainly in the domain of DMs, whereas comprehensive corrections have become increasingly popular in NDMs in recent years.

\begin{figure}[!ht]
	\begin{center}
		\begin{subfigure}{0.45\textwidth}
			\centering
			\begin{tikzpicture}[scale=0.8]
				\pie[polar, /tikz/every pin/.style={align=center},style={lines}, /tikz/nodes={text=black!80},text=pin,explode=0.1,rotate=100,color={teal!10,teal!30}, ]{45/DM, 55/NDM}
			\end{tikzpicture}
			\caption[]{{\small Proportions of deterministic and non-deterministic methods found in the reviewed literature.}}\label{fig:Method2}
		\end{subfigure}
		\vskip\baselineskip	
		\begin{subfigure}{0.40\textwidth}	
			\begin{tikzpicture}[scale=0.4]
				\pie[polar, radius=4, /tikz/every pin/.style={align=center},style={lines}, /tikz/nodes={text=black!80},text=pin,explode=0.1,rotate=160, color={teal!10,teal!20,teal!30,teal!40}]{10/Multiplicative, 17/Calibration, 28/Additive, 45/Comprehensive}
			\end{tikzpicture}
			\caption[]{{\small Distribution of non-deterministic methods found in the reviewed literature.}}\label{fig:comNDM}
		\end{subfigure}
		\hspace{1cm}
		\begin{subfigure}{0.40\textwidth}
			\begin{tikzpicture}[scale=0.4]
				\pie[polar, radius=4, /tikz/every pin/.style={align=center},style={lines}, /tikz/nodes={text=black!80},text=pin,explode=0.1,rotate=160, color={teal!10,teal!20,teal!30,teal!40}]{49/Multiplicative, 7/Space\\mapping, 32/Additive, 12/Comprehensive}
			\end{tikzpicture}
			\caption[]{{\small Distribution of deterministic methods found in the reviewed literature.}}\label{fig:comDa}
		\end{subfigure}
		\caption{Proportion of deterministic and non-deterministic methods utilized for constructing multi-fidelity surrogate models based on the literature. The chart also displays the distribution of the combination methods introduced in Section \ref{subsub:avail} within each category, deterministic and non-deterministic methods.}
		\label{fig:combination}
	\end{center}
\end{figure}

The landscape is evolving, with increasing emphasis on NDMs. This shift is likely due to the growing recognition of the importance of capturing complex uncertainties and correlations within MFM and the aid of increased computer power. The shift towards NDMs can also be attributed to the increasing demand for greater precision in modeling and simulation. As systems become complex, and the stakes increase, companies are more willing to invest in advanced UQ techniques. While previously, UQ was generally reserved for high-risk or high-cost applications, such as aerospace engineering or nuclear energy, the increasing need for precision in a broader range of industries has fueled this trend. Businesses now recognize the value of robust and reliable predictions for optimizing performance, ensuring safety, or reducing costs. \cite{dey2023uncertainty}.

The growing relevance of NDMs is further supported by publications from the 20th century primarily focused on the application of DMs. In contrast, the early years of the 21st century have witnessed a marked increase in the adoption of NDMs. Figure \ref{fig:Charts_year} presents histograms that show the distribution of scholarly articles on DMs and NDMs \footnote{The analysis considers literature up to 2016.}. Prominent examples of NDMs gaining traction include Kriging \cite{kleijnen2009kriging}, co-Kriging \cite{march2011gradient} and the Bayesian calibration models proposed by Kennedy and O'Hagan \cite{kennedy2001bayesian}.

\begin{figure}[!ht]
\vspace*{10pt}	\centering
	\begin{subfigure}[b]{0.45\textwidth}
		\centering
		\begin{tikzpicture}[font=\small,scale=0.7]
			\begin{axis}[
				ybar,
				bar width=10pt,
				xlabel={Year published},
				ylabel={Deterministic methods},
				ymin=0,
				ytick=\empty,
				xtick=data,
				axis x line=bottom,
				axis y line=left,
				enlarge x limits=0.1,
				symbolic x coords={1992,1995,1997,1999,2002,2005,2007,2010,2012,2015},
				xticklabel style={anchor=base,yshift=-\baselineskip, rotate=70},
				nodes near coords={\pgfmathprintnumber\pgfplotspointmeta\%}
				]
				\addplot[fill=teal!30, draw=none] coordinates {
					(1992,5)
					(1995,10)
					(1997,12)
					(1999,7)
					(2002,3)
					(2005,13)
					(2007,12)
					(2010,17)
					(2012,8)
					(2015,13)
					
				};
			\end{axis}
		\end{tikzpicture}
		\caption[]%
		{Despite the consistent prevalence of deterministic methods since 1990, the conducted analysis revealed the existence of two distinguishable peaks in the years 1996 and 2010.}
		\label{fig:Det}
	\end{subfigure}
	\quad
	\begin{subfigure}[b]{0.45\textwidth}
		\centering
		\begin{tikzpicture}[font=\small,scale=0.7]
			\begin{axis}[
				ybar,
				bar width=10pt,
				xlabel={Year published},
				ylabel={Non-deterministic methods},
				ymin=0,
				ytick=\empty,
				xtick=data,
				axis x line=bottom,
				axis y line=left,
				enlarge x limits=0.1,
				symbolic x coords={1983,1986,1990,1993,1997,2000,2004,2007,2011,2015},
				xticklabel style={anchor=base,yshift=-\baselineskip, rotate=70},
				nodes near coords={\pgfmathprintnumber\pgfplotspointmeta\%}
				]
				\addplot[fill=teal!30, draw=none] coordinates {
					(1983,2)
					(1986,0)
					(1990,2)
					(1993,0)
					(1997,0)
					(2000,4)
					(2004,8)
					(2007,13)
					(2011,27)
					(2015,44)
					
				};
			\end{axis}
		\end{tikzpicture}
		\caption[]%
		{Non-deterministic methods were scarce in the pre-21st century era. Nonetheless, there has been a notable rise in the application of non-deterministic methods since then.}
		\label{fig:Non_det}
	\end{subfigure}
	\caption{Frequency of publication over time for deterministic and non-deterministic methods.}
	\label{fig:Charts_year}
\end{figure}

\section{Reporting} \label{sec:Reporting}

\subsection{Open sourcing}

Open sourcing has emerged as a pivotal cornerstone and is considered a good practice for advancing scientific research, fostering transparency, reproducibility and community collaboration. It bridges the gap between theory and application by providing hands-on access to the tools and methods described in academic papers. For MFMs, open-sourcing the implementations not only aids fellow researchers in verifying claims but also accelerates the development and adoption of these tools across various domains. GitHub, GitLab and Bitbucket are common platforms for hosting open-source projects. Notable multi-fidelity publications that have generously shared their codebase, catalyzing further advancements and facilitating a more holistic understanding of their methodologies are Perdikaris et al. (2017) \cite{perdikaris2017nonlinear}, Costabal et al. (2019) \cite{costabal2019multi}, Eggensperger et al. (2021) \cite{eggensperger2021hpobench} and Pfisterer et al. (2022) \cite {pfisterer2022yahpo}.

\subsection{Benchmarking}

Benchmarking is also a good practice and provides a standardized basis for assessing various MFM methodologies. Commonly used benchmark functions have been instrumental in measuring the performance of these models. In multi-fidelity optimization, the following suite of benchmark functions has been frequently employed to evaluate and compare various methodologies.

The Forrester function, introduced by Forrester et al. (2008) \cite{forrester2007multi}, is a simple yet insightful test case, often used to introduce the basics of multi-fidelity optimization due to its tractability and interpretable low and high-fidelity versions. Additionally, the Branin function, known for its multi-modal nature, has been modified and employed in several multi-fidelity studies to challenge algorithms with its multiple optima. Dong et al. (2015) \cite{dong2015multi} introduced bi-fidelity versions of the Bohachevsky, Booth, Branin, Himmelblau and Six-hump Camelback functions. Toal (2015) \cite{toal2015some} introduced correlation-adjustable multi-fidelity versions of the Branin, Paciorek, Hartmann3 and Trid functions. Park et al. (2017) \cite{park2017remarks} defined the Hartman6 function for a multi-fidelity setting and used the multi-fidelity extension of the Borehole function, representing the flow of water through a borehole penetrating two aquifers. While the Rosenbrock function is ubiquitously utilized across optimization research, its applications in multi-fidelity studies, although less common, have provided insightful observations on the behavior of optimization algorithms in contending with its notorious banana-shaped valley \cite{wackers2023efficient}. Employing these benchmarks provides a standardized basis for algorithmic assessment and ensures a robust understanding of the method's strengths and weaknesses in various problem landscapes.

Surjanovic and Bingham \cite{surjanovic2013virtual} collected a collection of MATLAB/R implementations for some test functions and datasets found in the literature on emulation and prediction of computer experiments such as Borehole, Currin and the Park91 A and B functions. Meanwhile van Rijn and Schmitt \cite{van2020mf2} developed a Python package implementation (MF2) for the fixed multi-fidelity versions of Bohachevsky, Booth, Borehole, Branin, Currin, Forrester, Hartmann6, Himmelblau, Park91 A, Park91 B, and Six-Hump Camelback.

\subsection{Cost savings and accuracy report}

The utility of MFMs in terms of cost and time savings can vary depending on the problem being tackled. A pioneering study by Park et al. in 2017 \cite{park2017remarks} demonstrated that the use of MFMs can lead to up to 86\% cost savings without compromising the quality of the result, and accuracy improvements of up to 51\% can be achieved while maintaining the same cost. Despite this, other variables, such as the inherent structure of the problem or the type of SMs used, can significantly affect the extent of these savings. Several studies like Eldred et al. (2004) \cite{eldred2004second} and Peherstorfer et al. (2018a) \cite{peherstorfer2018survey} have delved into these complexities, indicating that a straightforward application of MFMs is rarely universally adequate.

To evaluate the effectiveness of MFMs, various metrics are commonly employed. The cross-validation error (CVE) and root mean squared error (RMSE) are prevalent choices. However, these metrics are not without their limitations. While CVE is generally effective for model comparison, it might not always identify the best model. Nonetheless, it is mainly beneficial for eliminating models that perform poorly \cite{park2017remarks, peherstorfer2016multifidelity}. RMSE is another metric that provides an objective numerical accuracy assessment but may fail to capture more complex model behavior.

To better understand MFM performance, it is helpful to evaluate them in comparison to their LFM and HFM counterparts in two aspects: accuracy and cost. The study by Peherstrofer et al. in 2016 \cite{peherstorfer2016multifidelity} provides valuable insights. They used RMSE to show how accurate MFMs are compared to LFMs and HFMs when the computation cost is fixed. They also showed how much time is saved using MFMs for a given level of accuracy compared to HFMs. The reader can find literature examples of MFM cost-efficiency in Table \ref{tab:MF_HF} of Appendix \ref{sec:tables}. It breaks down the cost differences between MFMs and HFMs and categorizes the information by application field, exposing the areas where MFMs have been most successful.

\begin{figure}[ht]
	\begin{center}
		\includegraphics[width=11cm]{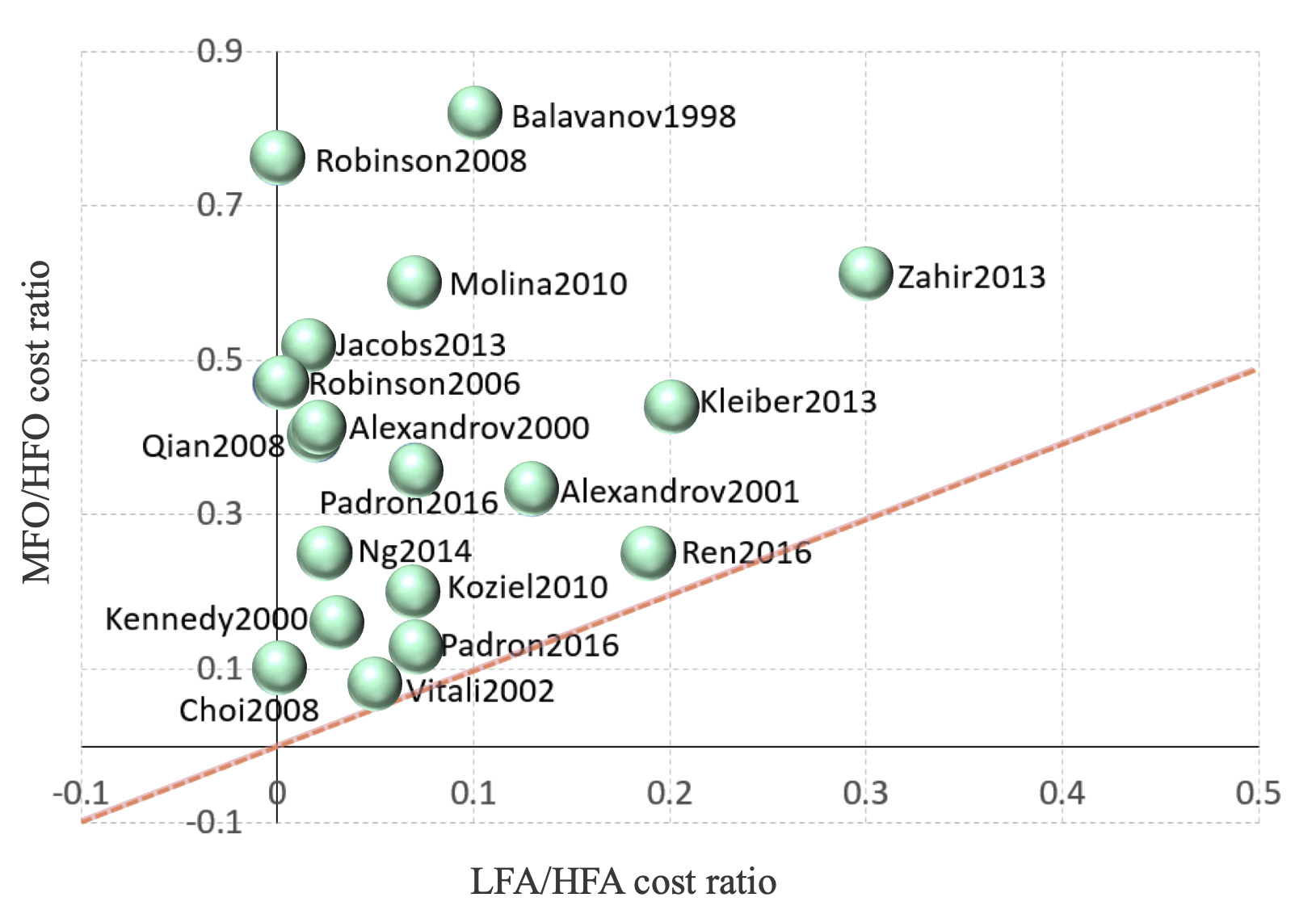}
		\caption{\label{fig:Cost} The cost ratio for a single analysis using a low-fidelity model compared to a high-fidelity model (known as the LFA/HFA cost ratio) is measured against the cost ratio for completing an optimization process with a multi-fidelity surrogate model versus a high-fidelity model (termed the MFO/HFO cost ratio). The dashed line in the graph serves as a threshold; points below this line suggest that using multi-fidelity models does not result in speed-ups.}
	\end{center}
\end{figure}

We also aimed to investigate whether there is a relationship between the cost-efficiency of LFMs and MFMs compared to HFMs in optimization tasks. In particular, we investigated the LFM/HFM cost ratio for a single analysis (LFA/HFA cost ratio) and the MFM/HFM cost ratio for a complete optimization process (MFO/HFO cost ratio). Figure \ref{fig:Cost} compiles data from 18 studies that explicitly reported these cost ratios out of 120 reviewed \footnote{The analysis considered papers published up to 2016.}. The 45-degree dashed line in the figure serves as a benchmark; data points below it do not indicate a cost advantage of using MFMs. Surprisingly, as the figure shows, our analysis revealed no significant correlation between the LFA/HFA and MFO/HFO cost ratios. This lack of correlation suggests that the cost-effectiveness of MFMs can be influenced by factors not considered in the figure. For instance, LFMs that are cheaper might also be less accurate, affecting the optimization speed. Additionally, the MFM's complexity could contribute to the total optimization cost.

This study recommends researchers using MFMs to offer a detailed cost, savings and accuracy breakdown concerning LFMs and HFMs. Such comprehensive data is sporadically present in existing literature but often buried, making it hard for readers to evaluate the merits and drawbacks of MFMs. Authors should present this critical information in a well-organized table to address this. This table should include (1) basic information such as the distinction between low- and high-fidelity models, details on any surrogate models and the methods for combining different fidelities; (2) cost-related metrics like comparisons of cost and accuracy between low-, high- and multi-fidelity models both at equivalent costs and for the same level of accuracy; and (3) cost-benefit analysis incorporating the time and resources invested in creating the MFM as well as comparisons with surrogate models if applicable. By adopting such a standardized presentation, researchers can offer a clearer picture of the efficacy and efficiency of MFMs in their studies.

As a case study, the research of Padr\'on et al. (2016) \cite{padron2016multi}  on airfoil optimization is an excellent example. Using RANS simulations as the HFM and Eulerian simulations as the LFM, their approach incorporated a stochastic polynomial chaos expansion as an SM. Future authors should consider emulating their comprehensive report. Table \ref{tab:1} captures critical metrics like the cost and error ratios between single fidelities (LF/HF) and between MFMs and HFMs (MF/HF), thereby furnishing an informative blueprint for subsequent research in the MFM domain.

\begin{table}[ht]
	\begin{center}
		\caption{The 2016 study by Padrón et al. \cite{padron2016multi}  serves as an exemplary guide for authors on how to report costs, savings and accuracy metrics.}
		\begin{tabular}{>{\centering\arraybackslash}m{1in} | >{\centering\arraybackslash}m{0.5in} | >{\centering\arraybackslash}m{1.5in} }
			\hline
			\textbf{Property}&  \textbf{Value}& \vphantom{a} \textbf{Comments} \\[5pt]
			\hline\noalign{\smallskip} \hline
			Cost LF/HF & 0.07 &  \vphantom{a} LF= Euler, HF= RANS \\ [5pt]\hline
			Error LF/HF & 0.18 & -\\ [5pt]\hline
			Cost MF/HF & 0.13 &  \vphantom{a} MF= 1 HF + 17 LF \\ [5pt]\hline
			Error MF/HF & 0.05 & - \\ [5pt]\hline
		\end{tabular}
		\label{tab:1}
	\end{center}
\end{table}

\section{Current trends} \label{sec:CurrentTrends}

MFMs have become an essential tool for optimization and UQ in numerous scientific and engineering fields in the last decade. Their recent evolution allowed more powerful algorithms to manage complex, high-dimensional challenges while giving dependable uncertainty estimates. A key advancement is the adoption of active learning, where strategically chosen data points from high-fidelity domains are used to update and refine LFMs. This iterative process allows for efficient design space exploration and leads to improved predictions. \cite{chaudhuri2021mfegra}. Deep learning, especially through architectures like CNNs and RNNs, has revolutionized the use of vast amounts of data. Previously, handling \textit{big data} meant shrinking it, often at the cost of valuable information \cite{zimmer2021auto}. Now, these neural network architectures help extract intricate features from detailed data, bridging gaps between fidelity levels and enhancing the accuracy of LFMs \cite{zhang2021multi, tao2019application}. Another significant trend in MFM research is the rise of transfer learning. By utilizing pre-trained neural networks that have learned representations from HFMs on related tasks, this approach allows efficient knowledge sharing between fidelity levels, achieving better performance while saving computational effort \cite{meng2020composite, chakraborty2021transfer, guo2022multi}. UQ in MFMs is another domain seeing progression. Bayesian neural networks (BNNs) are combined with MFMs to measure uncertainties across fidelity levels, using Bayesian techniques to update SMs based on new data continually \cite{meng2021multi}. Another promising avenue for UQ is anchoring-based neural networks (ABNNs). They use anchors or specific data clusters, each tied to a distinct neural network. This structure not only enhances uncertainty characterization but also provides insights into data regions or types, all while maintaining model performance \cite{anirudh2021delta}. As the field evolves, researchers are keenly exploring methods like transfer learning, fine-tuning and ensemble techniques to update these models efficiently without compromising their existing knowledge \cite{zhuang2020comprehensive, tajbakhsh2016convolutional, ardabili2020advances}.

\subsection{Interpretability challenges and physics-informed approaches}

ANNs have seen impressive advancements but come with unresolved challenges and questions. One primary concern is the interpretability of deep learning in MFMs. The inner workings of these models and meaningful explanations for their decisions are often elusive. Traditionally, data dimensionality was reduced before training SMs. However, the trend now leans towards tailoring data to make it suitable for platforms like PyTorch and TensorFlow. This shift can sometimes result in unsupervised pattern recognition, complicating interpretation or introducing bias. Addressing these challenges, Raissi et al. (2019) introduced physics-informed neural networks (PINNs) \cite{raissi2019physics}. PINNs present a unique approach: They learn solutions to partial differential equations directly through ANNs. These solutions are continuous, differentiable and respect fundamental physical laws, such as conservation of mass and momentum. Liu \& Wang (2019) \cite{liu2019multi} further suggested integrating physics-based constraints within ANNs. By doing so and using data from various fidelity levels, they aimed to boost model accuracy. These ANNs are designed such that their loss functions ensure these physical constraints are met, thereby fusing traditional physics insights with modern machine learning.

\subsection{Neural operators}

In recent years, the concept of neural operators has gained traction as a promising approach for solving differential equations, representing operators and generalizing functions from scattered data points. Unlike traditional ANNs that operate on fixed-size vectors, neural operators are designed to generalize across infinite-dimensional function spaces. However, a notable challenge with these models is their intensive computational demand, often requiring extensive amounts of high-fidelity data for training. Recent advancements have incorporated multi-fidelity data in training neural operators to mitigate this challenge. For instance, the works of Lu et al. (2022) \cite{lu2022multifidelity} and Howard et al. (2023) \cite{howard2022multifidelity} highlight strategies for leveraging both low-fidelity and high-fidelity data to train neural operators. Such an approach reduces the computational burden and enhances the generalization capabilities of the models.
Moreover, the research outlined in Ahmed and Stinis (2023) \cite{ahmed2023multifidelity} underscores the efficacy of multi-fidelity data in ensuring robust and efficient approximation of operators, especially in scenarios where high-fidelity data is either sparse or expensive to procure. Additionally, techniques like those presented in De et al. (2023) \cite{de2023bi} emphasize the cohesive integration of data from varying fidelities, ensuring that the neural operator captures the intricate dynamics and characteristics inherent in the high-fidelity data while also benefiting from the broader coverage and insights offered by the low-fidelity data. The confluence of neural operators and MFMs unveils novel research avenues poised to influence future computational methodologies in diverse domains significantly.

\subsection{Scalability advancements and challenges for MFMs}

Additionally, scalability and computational efficiency pose ongoing challenges, especially when dealing with large-scale optimization problems. The application of MFMs to complex systems often requires extensive computational resources due to the large number of simulations required for optimization or UQ. Modern tools like PyTorch Lightning have emerged to simplify the processing, training and scaling process for deep learning models, making it easier to utilize multi-GPU and TPU setups and automate repetitive code, thereby potentially aiding in MFM scalability \cite{falcon2019pytorchlightning}. Despite these tools, further research is necessary to develop scalable MFM frameworks, including exploring parallel analyses, GPU usage, exascale computing, distributed architectures, and quantum computing. By addressing these scalability and computational efficiency concerns, MFMs can be better equipped to tackle real-world applications.

\subsection{Future directions} \label{sec:FutureDirections}

As the field advances, the potential for deeper integration with active learning, neural operators and other methodologies is evident. The push for interpretability in scientific deep learning-based MFMs will likely remain at the forefront of research concerns. PINNs represent a promising direction, hinting at future MFMs that meld physics with data-driven insights. There is a palpable momentum towards an integrated modeling framework where transfer learning, Bayesian methods, and UQ come together. With these developments on the horizon, the next phase in MFM promises revolutionary breakthroughs.

\section{Conclusion} \label{sec:Conclusion}

This review has examined multi-fidelity models' growing importance and versatility in scientific and engineering domains, emphasizing their ability to optimize computational resources while maintaining high predictive accuracy. The paper highlighted distinctive characteristics and a wide range of applications through a classification scheme, notably identifying optimization and uncertainty quantification as the most prevalent use cases, accounting for approximately 90\% of the reviewed publications.

The study revealed two predominant fidelity-management strategies: one that integrates different fidelities into a surrogate model and another that employs a hierarchical utilization of models based on specific criteria. Our analysis indicated that the former strategy was more prevalent, representing approximately 70\% of the reviewed literature. Notably, recent advancements in computational power have led to increased hierarchical contributions in the past few years. In addition, the past two decades have shown a noticeable shift toward non-deterministic methods due to their ability to offer uncertainty estimates.

Despite the promise of cost efficiency, this study found no direct correlation between cost ratios of low- and high-fidelity analyses and those of multi-fidelity vs. high-fidelity optimizations, indicating the case-specific nature of these problems. The review also called for standardized reporting metrics to improve the field's understanding and evaluation.

Recent advancements in artificial neural networks have notably influenced multi-fidelity modeling but have also raised concerns about interpretability and scalability. These challenges will be critical in determining the broader adoption of these methodologies in complex optimization and uncertainty quantification tasks. Open-sourcing and benchmarking approaches are discussed in this work to promote transparency, reproducibility and community engagement. The review concluded by highlighting emerging research trends, and it serves as a foundational reference for ongoing and future work in the evolving field of multi-fidelity modeling.

\appendix

\section{Nomenclature}\phantom{text} \label{sec:Nomenclature} 

The following definitions are organized to facilitate the comprehension of the document.

\subsection{Basic terminology}
\begin{description}
	\item[Analysis] A single evaluation of a model or process. \textit{See also: Data, Datum, Simulation}.
	\item[Data] The outcome of multiple analyses. \textit{See also: Datum}.
	\item[Datum] The outcome of a single analysis. \textit{See also: Data}.
	\item[Experiment] A real-world test, as opposed to a simulation.
	\item[Fidelity] Level of accuracy in a model or process.
	\item[Model] A mathematical representation of a physical phenomenon. \textit{See also: HFM, LFM, MFM}.
	\item[Point] The value a variable can take. It serves as the input for an analysis.
	\item[Response] Exchangeable with \textit{Analysis}.
\end{description}

\subsection{Types of models and analyses}
\begin{description}
	\item[HFA] High-Fidelity Analysis. A single evaluation using a high-fidelity model. \textit{See also: HFM}.
	\item[HFM] High-Fidelity Model. A model that estimates outputs with high accuracy \cite{peherstorfer2018survey}. \textit{See also: HFA}.
	\item[LFA] Low-Fidelity Analysis. A single evaluation using a low-fidelity model. \textit{See also: LFM}.
	\item[LFM] Low-Fidelity Model. A model that estimates outputs with lower accuracy, typically for lower costs \cite{peherstorfer2018survey}. \textit{See also: LFA}.
	\item[MFM] Multi-Fidelity Model. A model combining information from multiple models of different accuracies. \textit{See also: MFHM, MFSM}.
	\item[SM] Surrogate Model. Algebraic approximation fitted to available data points.
	\item[Data Point / Sampling Point] Information used for training a surrogate model.
	\item[Data Fit] The process of using available data points to construct a surrogate model.
\end{description}

\subsection{Advanced concepts}
\begin{description}
	\item[DM] Deterministic Method. Basis functions are assumed, and their coefficients are found by minimizing discrepancies between data and functions.
	\item[HFSM] High-Fidelity Surrogate Model. Constructed using a high-fidelity model and may also be treated as such.
	\item[LFSM] Low-Fidelity Surrogate Model. Constructed using data points from a low-fidelity model.
	\item[MFHM] Multi-Fidelity Hierarchical Model. No multi-fidelity surrogate model is constructed; fidelity is chosen by a criterion\footnote{Multi-fidelity surrogate models and multi-fidelity hierarchical models are termed as multi-fidelity management strategies by Peherstofer et al. (2018a) \cite{peherstorfer2018survey}}.
	\item[MFSM] Multi-Fidelity Surrogate Model. Constructed using multiple models of different accuracies. These models can be surrogate models themselves.
	\item[NDM] Non-Deterministic Method. Assumes the function or its coefficients are uncertain; uses samples to reduce uncertainty.
	\item[Outer-loop application] Involves computational applications that form outer\break loops around a model, such as optimization and uncertainty propagation \cite{peherstorfer2018survey}.
\end{description}

\section{Design of experiments} \label{sec:DesignOfExperiments}

\subsection{Traditional sampling methods}
\label{subsec:trad_sampl}
\begin{figure}[ht]
	\centering
	\begin{subfigure}[b]{0.475\textwidth}
		\centering
		\includegraphics[width=2in]{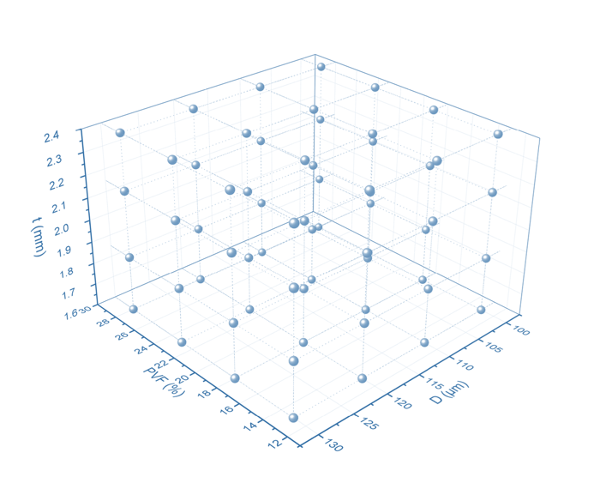}
		\caption{Full factorial design with 3 factors and 4 levels.}
		\label{fig:FFD}
	\end{subfigure}
	\begin{subfigure}[b]{0.475\textwidth}
		\centering
		\includegraphics[width=1.5in]{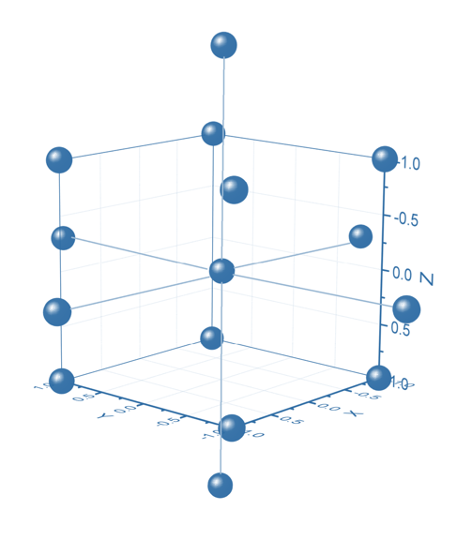}
		\caption{Central composite design with 3 factors and 5 levels.}
		\label{fig:CCD}
	\end{subfigure}
	\hfill
	\caption{Examples of sampling strategies.}
	\label{fig:sampling_strategies}
\end{figure}

The construction of SMs needs a suitable sampling technique for generating a representative set of data points. The choice of the sampling methodology is crucial to the SM's accuracy, as illustrated in the work of Dribusch et al. (2010) \cite{dribusch2010multifidelity}. Grid-based sampling methods, such as full factorial design (FFD), entail the sampling of each variable (factor) at a fixed number of levels and are generally used for low-dimensional problems (typically involving less than three variables), as depicted in Figure \ref{fig:FFD}. As shown in the work of Fern\'andez-Godino et al. (2016) \cite{fernandez2016anomaly}. On the other hand, the central composite design (CCD) method extends the two-level FFD by augmenting it with the minimum number of points required for each variable to provide three levels, enabling the fitting of a quadratic polynomial. This method is frequently employed for problems with three to six design variables, as shown in Figure \ref{fig:CCD}. For higher dimensional problems, an option is to use only a subset of the CCD vertices in an approach called small composite design (SCD), as described in the work of Myers and Montgomery (1995), pp. 351-355 \cite{montgomery1995response}. It should be noted that FFD, CCD and SCD are inflexible about the number of sampling points and domain shape.

\subsection{Optimality criteria-based designs}
\label{subsec:opt_crit}
\begin{figure}[ht]
	\begin{center}
		\includegraphics[width=12cm]{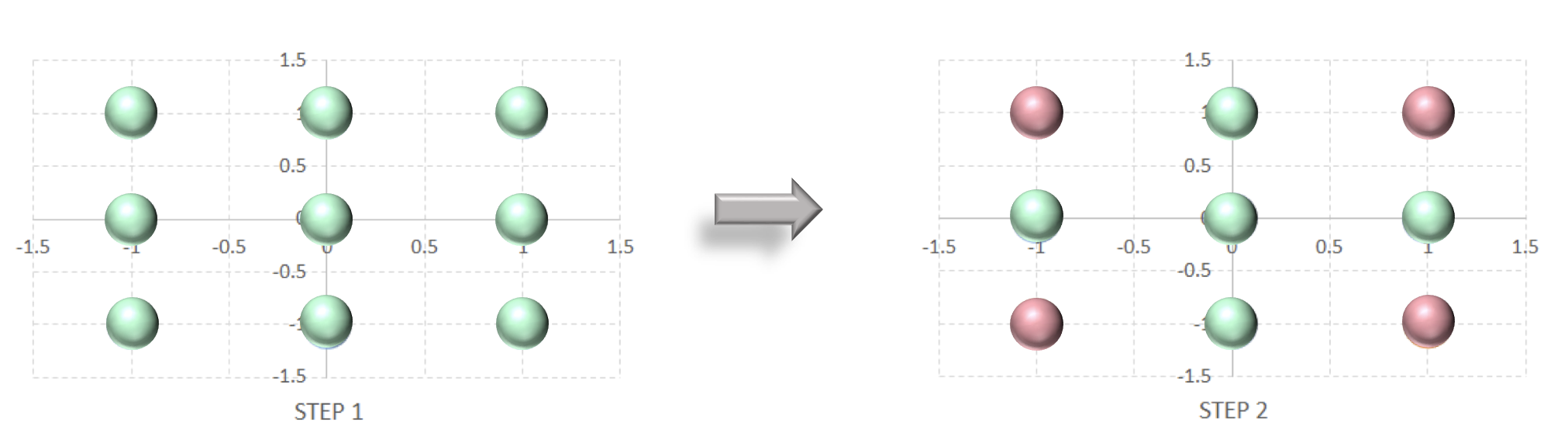}
		\caption{\label{fig:Strategy1} Example of a nested sampling design, where the teal-colored bubbles represent LFM points, and the pink-colored bubbles represent HFM points selected using the D-optimal design criterion.}
	\end{center}
\end{figure}

Designs of experiments that allow for any number of samples are frequently based on an optimality criterion. For instance, the D-optimal design approach \cite{de1995d} selects a subset of a grid in any domain shape by minimizing the determinant of the Fisher information matrix \cite{mentre1997optimal}, which reduces the impact of noise on the fitted polynomial and often results in a significant fraction of points at the boundary of the domain. Figure \ref{fig:Strategy1} illustrates the application of the D-optimal criterion in a nested sampling design for MFMs.

\subsection{Space-filling methods}
\label{sec:space_fill}

Space-filling methods, such as Monte Carlo and Latin hypercube sampling, are more commonly employed when the noise in the data is not a concern and uniformly distributed points are desired. In such cases, using an optimality criterion method to sample near the domain boundaries is preferable. The most popular variant of Latin hypercube sampling attempts to maximize the minimum distance between points, also known as the minimax criterion \cite{johnson1990minimax}, to promote uniformity.

\subsection{Considerations for multi-fidelity models}
\label{subsec:mfm_consider}

When dealing with MFSMs, the relationship between the sampling points of the LFMs and HFMs is an additional issue. A nested design sampling strategy generates HFM points as a subset of LFM points or LFM points as a superset of HFM points. It was initially developed as a space-filling method for generating additional datasets to complement existing ones using a criterion. For example, Jin et al. (2005) \cite{jin2005efficient} used three optimality criteria: minimax distance criterion, entropy criterion and centered $L_{2}$ discrepancy criterion.

Combining the original and additional points results in the sampling points for an LFSM, with the additional subset reserved for the HFSM \cite{park2017remarks}. Haaland \& Quian (2010) proposed nested design sampling for categorical and mixed factors \cite{haaland2010approach}. Zheng et al. (2015) compared the effects of nested and non-nested design sampling on modeling accuracy \cite{zheng2015difference}.

Incorporating the HFM points as a subset of LFM points simplifies the parameter estimation process for discrepancy function-based methods. However, if the HFM points are not a subset of the LFM points, the parameter estimation of the discrepancy function relies on the LFSM parameter determination. For instance, the co-Kriging method utilizes GP to model uncertainties for the LFSM and the discrepancy function. If the design of experiments satisfies the nested sampling condition, the parameters of each GP model can be estimated independently.

Although not all MFSMs adhere to this approach, some, such as those used for combining computer simulation results, allow for the control of input settings and the possibility of a nested set. Multiple options exist for nested designs, including generating the LFM design of experiments first and then selecting a subset using a specific criterion. This approach was used in Balabanov et al. (1998), where 2107 points were generated in a 29-dimensional space using SCD for LFM sampling points. Then, 101 sampling points were selected using the D-optimality criterion \cite{balabanov1998multifidelity}.

\subsubsection{Alternative approaches}
\label{subsubsec:alt_approach}

\begin{figure}[ht]
	\begin{center}
		\includegraphics[width=12cm]{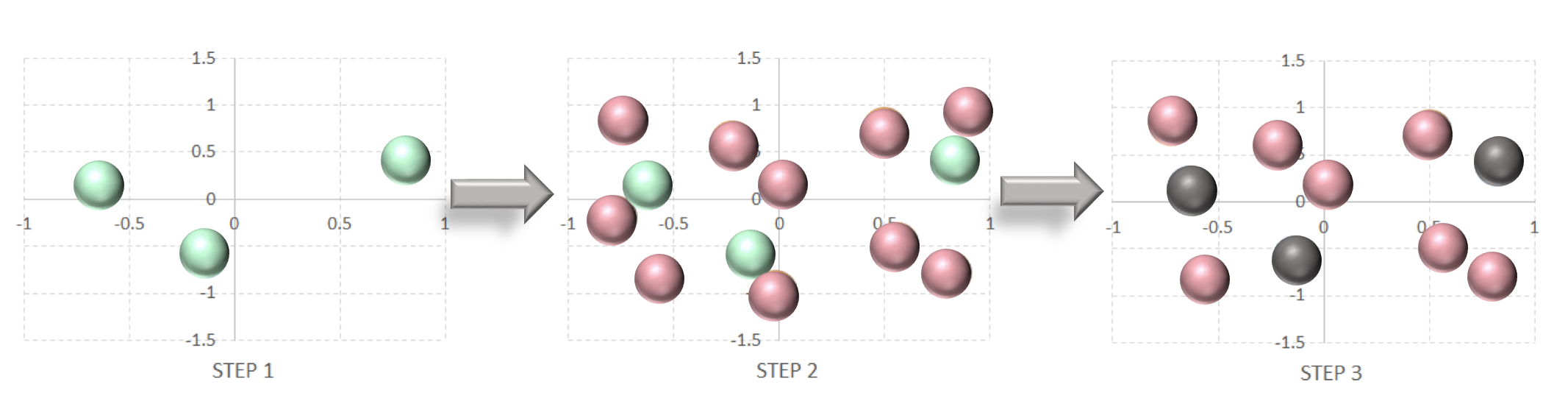}
		\caption{\label{fig:Sampling_proc} Nearest neighbor sampling. High-fidelity model points (teal bubbles) and low-fidelity model points (pink bubbles) are sampled independently, and then the low-fidelity model nearest neighbor point to each high-fidelity model point is moved on top of it (black bubbles)}
	\end{center}
\end{figure}

In Le Gratiet's study from 2013 \cite{le2013multi}, the generations of sampling points from both the LFM and HFM were conducted independently. Subsequently, the nearest LFM point to each HFM point was moved onto their corresponding nearest neighbor, as depicted in Figure \ref{fig:Sampling_proc}. This technique is commonly known as nearest-neighbor sampling.

\subsubsection{Adaptive sampling strategies}
\label{subsubsec:adaptive_sampl}

Adaptive sampling methods are common SM strategies that minimize the number of simulations required to construct a model to a specified accuracy by utilizing efficient interpolation and sampling techniques. For example, in a study by Mackman et al. in 2013 \cite{mackman2013comparison}, two adaptive sampling strategies for generating SMs based on Kriging and radial basis function interpolation were compared. The authors found that both strategies outperformed traditional space-filling methods.

Another option is the optimal placement of LFM design points for local searches in an iterative fashion, as evidenced in studies by Robinson et al. in 2006 \cite{robinson2006strategies} and Raissi and Seshaiyer in 2014 \cite{raissi2014multi}.

\section{Surrogate models} \label{sec:SurrogateModels}

\subsection{Introduction to surrogate models}
A surrogate model serves as a replacement for complex, costly or time-consuming simulations or experiments. They are also known as a metamodel, emulator, projection-based model, reduced-order method or data-fit model. Surrogate models (SMs) are constructed using available data that capture the relationship between input variables and output quantities of interest. Typically, SMs are algebraic functions developed by fitting a limited data set from expensive simulations or experiments, accelerating the prediction of the quantities of interest. The accuracy of an SM depends on various factors, such as the experimental design for selecting data points, the domain size, the dimensions of the input and output, the accuracy at the data points, and the number of samples available for constructing the SM \cite{simpson2001metamodels}.

\subsection{Application in multi-fidelity models}
In the domain of MFMs, SMs can be developed separately for each fidelity and utilized hierarchically within an MFHM approach, as demonstrated by Nelson et al. (2007) \cite{nelson2007multi} and Koziel and Leifsson (2013) \cite{koziel2013multi}. Alternatively, a unified MFSM can incorporate data from diverse physical models or experiments, even when those sources are already SMs, as shown in the studies by Giunta et al. (1995) \cite{giunta1995variable}, Qian et al. (2008) \cite{qian2008bayesian} and Padr'on et al. (2016) \cite{padron2016multi}.

\subsection{Types of surrogate models}
\subsubsection{Basis function regression}
Basis function regression, a generalization of response surface modeling \cite{box1992experimental}, is one of the oldest and most used forms of SM in engineering design. Its training process only requires solving a set of linear algebraic equations. It assumes that the functional behavior, such as a second-order polynomial, is accurate, but the output is subject to noise. In the context of MFMs, basis function regression has been extensively used, as evidenced by numerous research papers, including Chang et al., 1993 \cite{chang1993sensitivity}, Burgee et al. (1994) \cite{burgee1994parallel}, Venkatarman et al. (1998) \cite{venkataraman1998design}, Balabanov et al. (1998) \cite{balabanov1998multifidelity}, Balabanov et al. (1999) \cite{balabanov1999reasonable}, Mason et al. (1998) \cite{mason1998variable}, Vitali et al. (1998) \cite{vitali1998correction}, Knill et al. (1999) \cite{knill1999response}, Vitali et al. (2002) \cite{vitali2002multi}, Umakant et al. (2007) \cite{umakant2007ranking}, Venkatarman et al. (2006) \cite{venkataraman2006reliability}, Choi et al. (2008) \cite{choi2008multifidelity}, Sharma et al. (2008) \cite{sharma2008multi}, Sharma et al. (2009) \cite{sharma2009multi}, Sun et al. (2010) \cite{sun2010two}, Goldsmith et al. (2011) \cite{goldsmith2011effects} and Chen et al. (2015) \cite{chen2015effective}.

\subsubsection{Polynomial Chaos Expansion}
Polynomial chaos expansion (PCE) has gained popularity in the 21st century as a method for analyzing aleatory uncertainties using probabilistic approaches in UQ \cite{ghanem1990polynomial, sakamoto2002polynomial, xiu2002wiener, xiu2003modeling}. In PCE, a polynomial function is constructed to map uncertain inputs to the outputs of interest, and the statistics of the outputs are approximated. The chaos coefficients are estimated by projecting the system onto a set of basis functions (e.g., Hermite, Legendre Jacobi). PCE has been applied in the context of MFM in several studies, such as Eldred (2009) \cite{eldred2009recent}, Ng and Eldred (2012) \cite{ng2012multifidelity}, Padr\'on et al. (2014) \cite{padron2014multi}, Padr\'on et al. (2016) \cite{padron2016multi} and Absi and Mahadevan (2016) \cite{absi2016multi}.

\subsubsection{Kriging}
Kriging has gained significant popularity as an SM, particularly for applications in MFSM. This emergence may be attributed to its uncertainty structure, conducive to probabilistic MFSM, as discussed in Section \ref{sec:CombiningFidelities}. Kriging estimates the value of a function as the sum of a trend function (e.g., polynomial) representing low-frequency variation and a systematic departure representing high-frequency variation components \cite{queipo2005surrogate}. Unlike basis function regression, Kriging assumes that the data points response is correct, but the functional behavior is uncertain. Various studies have employed Kriging methods in the context of MFM, such as those by Leary et al. (2003) \cite{leary2003method}, Forrester et al. (2007) \cite{forrester2007multi}, Goh et al. (2013) \cite{goh2013prediction}, Huang et al. (2013, 2014) \cite{huang2013research}, Biehler et al. (2015) \cite{biehler2015towards} and Fidkowski et al. (2014) \cite{fidkowski2014quantifying}.

\subsubsection{Co-Kriging}
Co-Kriging, which extends Kriging to fuse multiple fidelity models, is regarded as a technique to construct an HFM approximation enhanced by data from LFMs. Relevant studies on co-Kriging methods can be found in Chung and Alonso (2002) \cite{chung2002design}, Forrester et al. (2008) \cite{forrester2008engineering}, Yamazaki et al. (2010) \cite{yamazaki2010design} and Han et al. (2013) \cite{han2013improving}. Laurenceau and Sagaut (2008) \cite{laurenceau2008building} compared the performance of Kriging and co-Kriging.

\subsubsection{Moving least squares}

Moving least squares (MLS) is a technique first proposed by Lancaster and Salkauskas in 1981 \cite{lancaster1981surfaces} and extensively discussed by Levin in 1998 \cite{levin1998approximation}. MLS is an improvement over the weighted least-squares (WLS) method, which Aitken introduced in 1935 \cite{aitken1936iv}. WLS recognizes that not all design points are equally important in estimating the polynomial coefficients. Thus, a WLS model is a straightforward polynomial, but with the fit biased toward points with a higher weighting. In an MLS model, the weightings are varied depending on the distance between the point to be predicted and each observed data point. Several studies have implemented MLS in MFM for various applications such as multi-point optimization \cite{toropov1999multipoint}, multi-fidelity analysis \cite{zadeh2002multi,zadeh2005use,berci2011multifidelity} and aerodynamic shape optimization \cite{sun2011multi}.

\subsubsection{Proper orthogonal decomposition}

Traditional SMs typically predict scalar responses, whereas some nontraditional SMs, such as proper orthogonal decomposition or POD, are used to obtain the entire solution field to a partial differential equation. Toal in 2014 \cite{toal2014potential}, Roderick et al. in 2014 \cite{roderick2014proper} and Mifsud et al. in 2016 \cite{mifsud2016variable} have explored the MFM proper orthogonal decomposition method in fluid mechanics.

\subsection{Advancements in surrogate modeling}
\subsubsection{Support vector machines}

Recent advances in computer power have paved the way for using more sophisticated and computationally expensive SMs, including support vector machines (SVM) and artificial neural networks (ANNs). These advanced SMs excel in handling highly non-linear and high-dimensional functions with the caveat that they require a large amount of data, and their training is computationally intensive.

SVMs widespread in the scientific community arrived in the late 1990s starting with the work of Hearst et al. (1998) \cite{hearst1998support}. SVMs are a type of supervised learning algorithm that can be used for both classification and regression tasks. They are based on finding an optimal hyperplane that separates data points belonging to different classes with the maximum margin. SVMs use kernel functions to map data into a higher-dimensional space where linear separation is possible.

\subsubsection{Artificial neural networks}
Although the concept of artificial neural networks or ANNs was initially introduced in the 1980s by Rumelhart and colleagues \cite{rumelhart1986learning}, they have only recently gained popularity due to the increased computational power that allows the handling of the massive amount of data they require. It was not until the late 2010s that they had their explosive emergence in the engineering world. This delay could be attributed to the scientific community finding it hard to interpret their architecture, referring to them as \textit{black boxes}. ANNs are constructed with a layered architecture consisting of individual neurons that calculate a weighted sum of input values. Note that the absence of activation functions at the output of an ANN layer would result in purely linear operations, rendering the network incapable of effectively modeling non-linearities. Activation functions are a solution to this problem and are employed to generate non-linear transformations of the output of these neurons. Minisci and Vasile (2013) \cite{minisci2013robust} provide an example of ANNs' application in MFM, where it is used in the optimization process to correct the aerodynamic forces in the simplified LFM using CFD simulations as the HFM. The LFM generates samples globally across the design parameter range, while the HFM is used to refine the ANN locally in later optimization stages. Liu and Wang (2019) \cite{liu2019multi} propose physics-constrained neural networks (PCNN) to improve training efficiency by reducing data requirements and incorporating physical knowledge as constraints. Their MFM ANN model demonstrates improved prediction accuracy compared to classical ANNs. Furthermore, the optimization framework maintains accuracy while reducing data needs. PCNNs benefit high-dimensional problems and real-world engineering applications with data sparsity challenges.

\section{Toy problems} \phantom{text}  \label{sec:ToyProblems}

By presenting these examples, the intention is to allow the reader to train simple MFMs and use them as a starting point for further in-depth study and exploration. Each of the following examples can be found in the review of multi-fidelity models' \href{https://github.com/mgisellef/ReviewOfMultiFidelityModels_ToyProblems}{GitHub repository}.

\subsection{Example 1: Additive and multiplicative corrections}

\begin{figure}[!ht]
	\begin{center}
		\includegraphics[width=8cm]{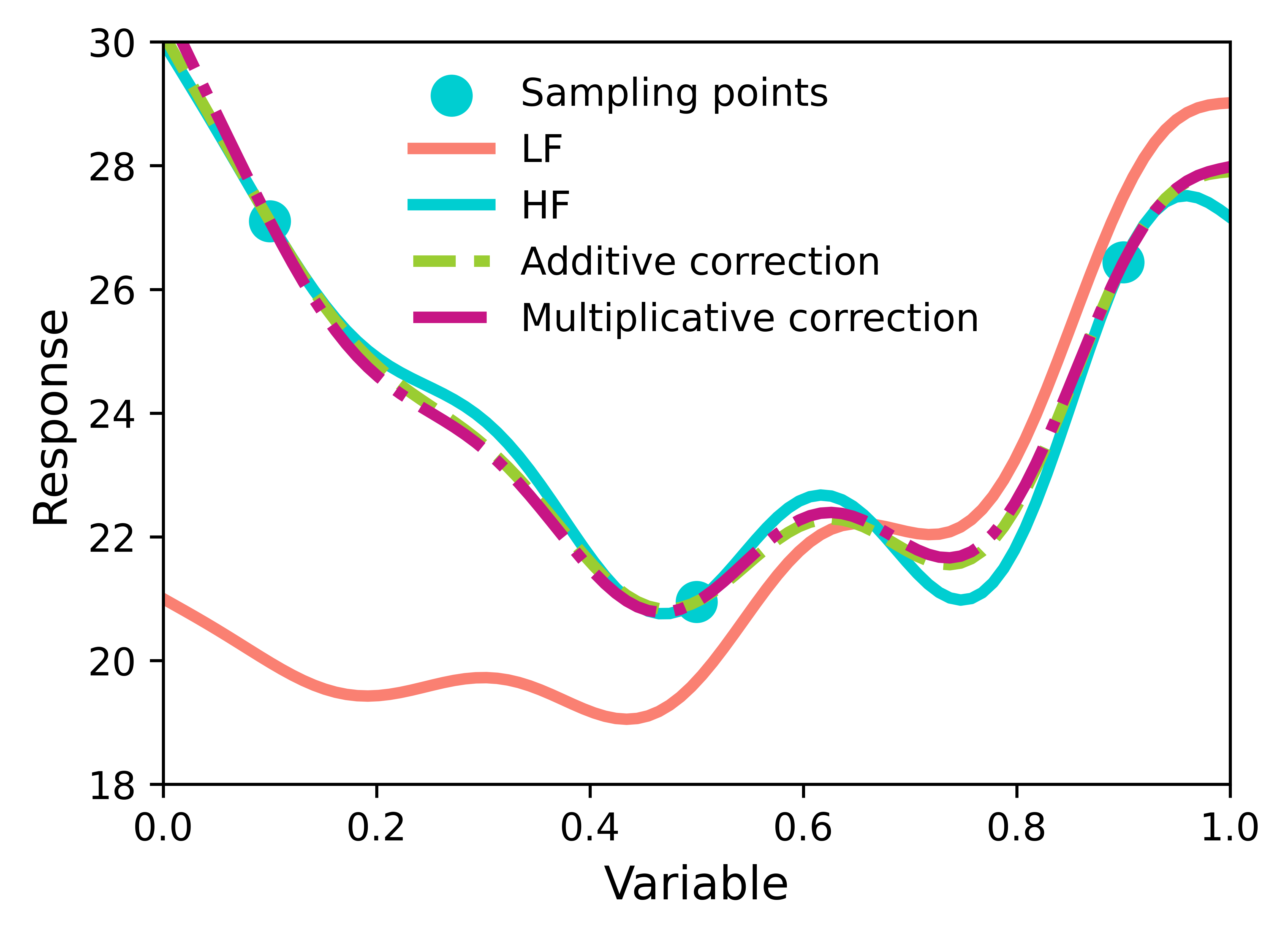}
		\caption{\label{fig:correction} One-dimensional analytic example illustrating the performance of additive and multiplicative correction approach.}
	\end{center}
\end{figure}

To illustrate the concepts of additive and multiplicative corrections, a simple algebraic one-dimen\-sional problem is presented. Consider two analytical functions, $y_{LF}$ and $y_{HF}$, given by Equations \eqref{LFfunc} and \eqref{HFfunc}, respectively:
\begin{equation}\label{HFfunc}
	y_{HF}(x)= 2x \sin(20x + 2) + 10 e^x + 20 (x-1)^2
\end{equation}

\noindent and
\begin{equation}\label{LFfunc}
	y_{LF}(x) = Ay_{HF}(x) + B(x-0.5) + C
\end{equation}

\noindent where $x \in [0,1]$, $y_{HF}$ is the HFM, and $y_{LF}$ represents the LFM, and  $A= 0.7$, $B= 10$ and $C= 5$. LFM and HFM may be unavailable in practice, and SMs are used instead. For this toy problem, it is assumed that the functions are accessible, but due to cost constraints, only three samples from the HFM can be afforded. Evenly distributed samples are chosen, specifically $[x_1,x_2,x_3]=[0.1,0.5,0.9]$, from two models, and the objective is to estimate the output of the HFM while minimizing expenses using LFM predictions. Additive and multiplicative corrections are introduced to achieve this, as shown in Figure \ref{fig:correction}. The ratio $y_{HF}$/$y_{LF}$ or the differences $y_{HF}$-$y_{LF}$ at the sampling points $x_i$ are used to fit the multiplicative or additive corrections, respectively. In simpler terms, the sampling points $[y_{HF}(x_i)/y_{LF}(x_i)]$ for $i \in {1,2,3}$ are used to fit the multiplicative correction $\rho(x)$, whereas the sampling points $[y_{HF}(x_i)-y_{LF}(x_i)]$ for $i \in {1,2,3}$ are used to fit the additive correction $\delta$. The functions $\rho(x)$ and $\delta(x)$ (Eqs. \eqref{sum} and \eqref{division}, respectively) are obtained using basis function regression by fitting a linear combination of second-order polynomial basis functions where the polynomial coefficients are optimized to minimize the residuals (\cite{zhang2018multifidelity,fernandez2019linear}). Figure \ref{fig:correction} displays the original HFM and LFM within the interval [1,2], alongside the additive and multiplicative corrections and the sample locations.

Although the performance appears to be similar in this particular case, note that for more complex systems, there can be significant variations in performance that may favor one approach over the other. The reader interested in replicating these results is encouraged to access example 1 \href{https://github.com/mgisellef/ReviewOfMultiFidelityModels_ToyProblems/blob/main/Comprehensive_Corrections.ipynb}{notebook}.

\subsection{Example 2: Comprehensive correction}

\begin{figure}[!ht]
	\begin{center}
		\includegraphics[width=8cm]{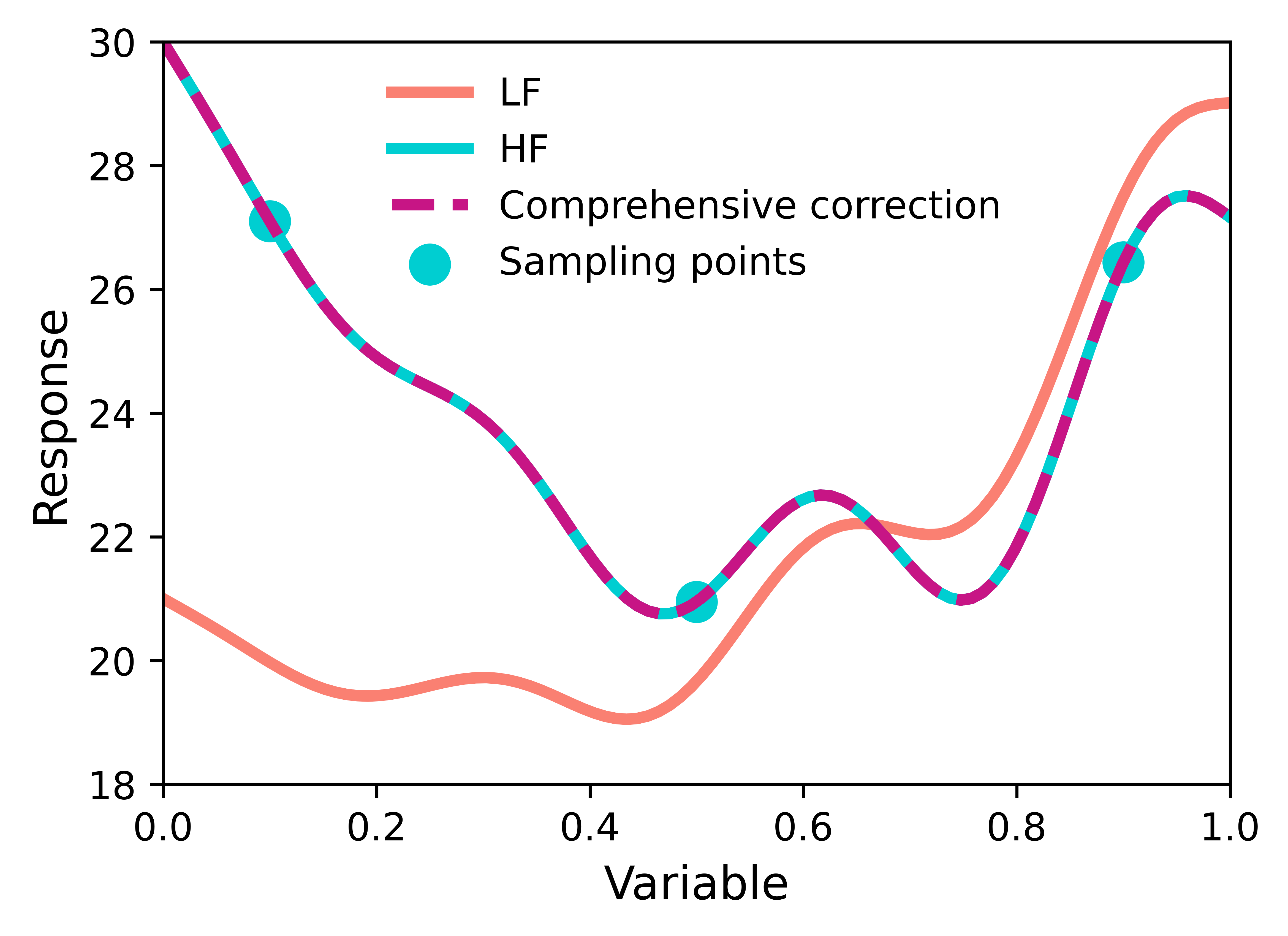}
		\caption{\label{fig:comp_correction} One-dimensional analytic example illustrating the concept of comprehensive correction \cite{zhang2018multifidelity}.}
	\end{center}
\end{figure}

The method involves creating a composite model that combines information from both LF and HF models to improve predictive accuracy. In this example, we illustrate the concept of comprehensive correction within a one-dimensional analytic setting, as depicted in Figure \ref{fig:comp_correction}. The comprehensive approach used is the one proposed by Zhang et al. (2018) \cite{zhang2018multifidelity}.

\begin{figure}[!ht]
	\centering
	\begin{subfigure}{0.25\textwidth}
		\centering
		\includegraphics[width=4.5cm]{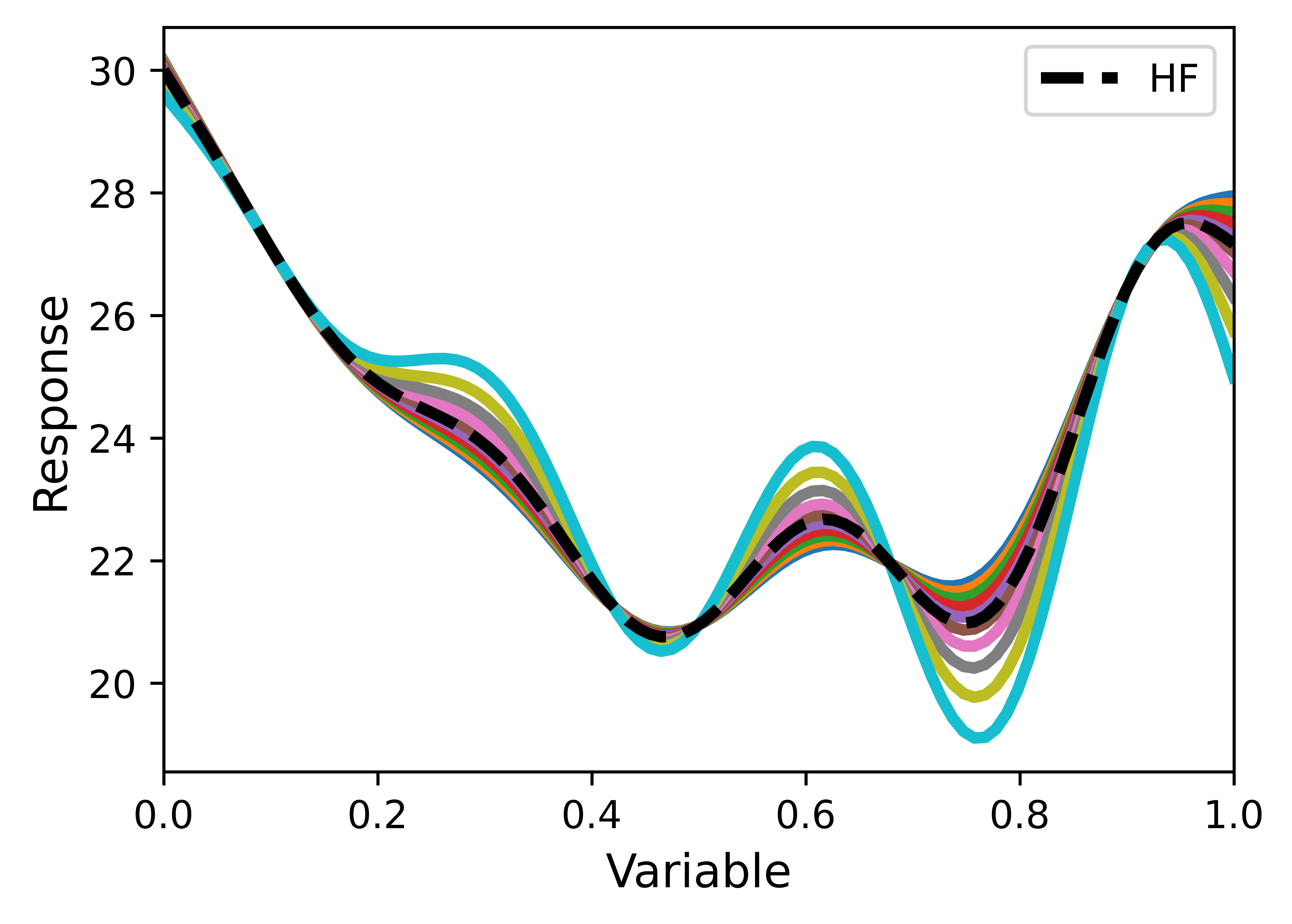}
		\caption{A.}\label{fig:A}
	\end{subfigure}
	\hspace{2cm}
	\begin{subfigure}{0.25\textwidth}
		\centering
		\includegraphics[width=4.5cm]{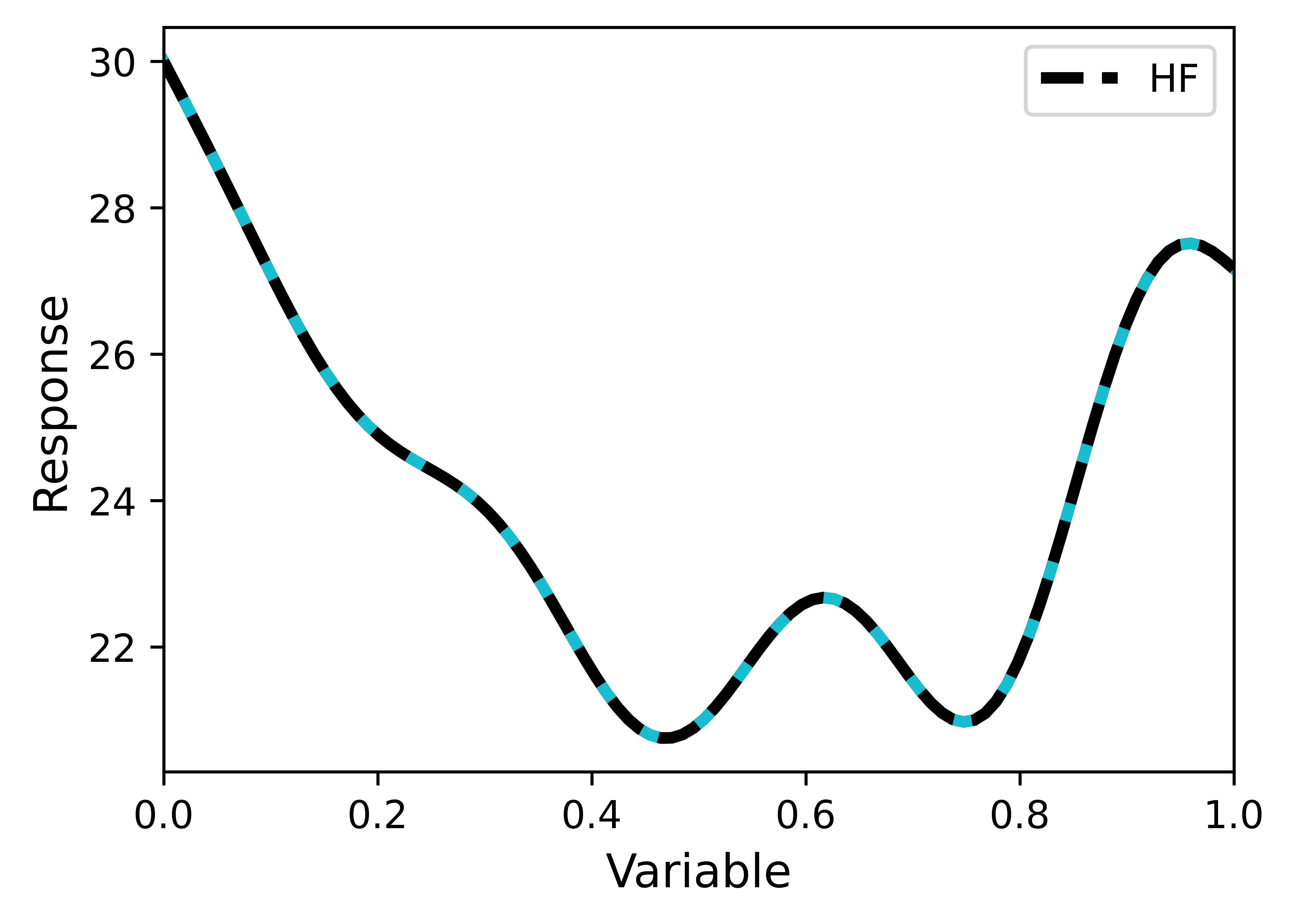}
		\caption{B.}\label{fig:B}
	\end{subfigure}
	\hspace{2cm}
	\begin{subfigure}{0.25\textwidth}
		\centering
		\includegraphics[width=4.5cm]{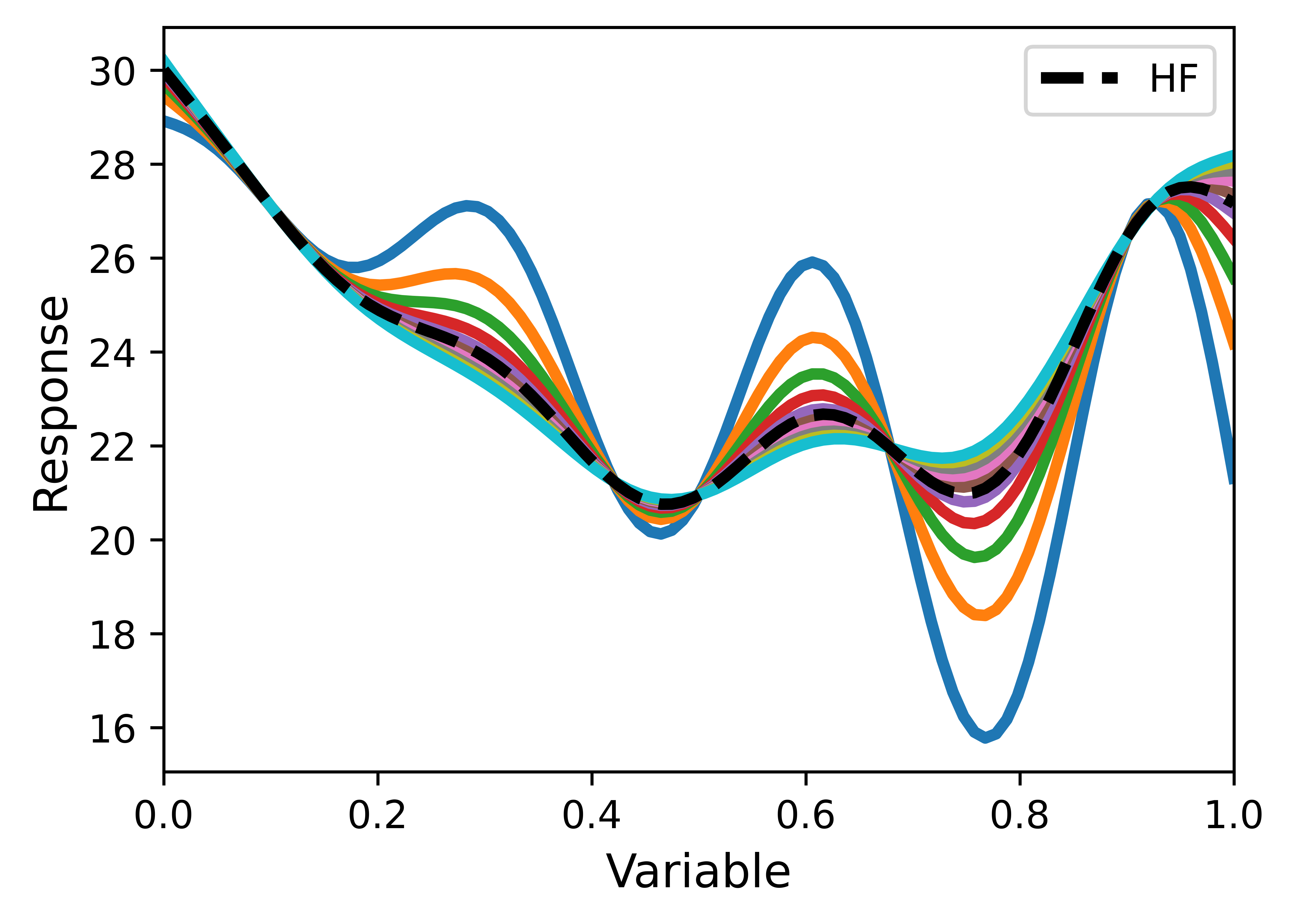}
		\caption{C.}\label{fig:C}
	\end{subfigure}
	\hspace{1cm}
	\caption{Effect of the LFM constants A, B, C on co-Kriging performance. The variables were modified one at a time, setting the ones not being considered to their default value. \label{fig:constants}}
\end{figure}

The study delves into the coefficients of the functions represented by Figure \ref{fig:constants}, which have been previously defined. This example demonstrates the application of comprehensive correction to refine the LFM using HF data, showcasing the power of multi-fidelity modeling in improving predictive accuracy. The reader interested in replicating these results is encouraged to access the example 2 \href{https://github.com/mgisellef/ReviewOfMultiFidelityModels_ToyProblems/blob/main/Additive_Multiplicative_Corrections.ipynb}{notebook}.

\subsection{Example 3: Multiple choice}

In building an MFSM, let us consider a scenario with limited availability of 20 data points from the HFM and 200 from the LFM. The development involves two key steps.

\vspace*{4pt}\noindent\textbf{Step 1.}
Construct an SM approximating the discrepancy or ratio between the HFM and LFM using the 20 shared data points.

\vspace*{4pt}\noindent\textbf{Step 2.}
Two options are available:
\begin{itemize}
	\item \textit{Option A}: Build an LFSM with the 200 LFM points and integrate it with the SM from Step 1, either through addition or multiplication.
	\item \textit{Option B}: Use the SM from Step 1 to estimate the discrepancy or ratio for the remaining 180 LFM-only points. Then, adjust these LFM points to yield estimated HFM values.
\end{itemize}

After these steps, both actual and estimated HFM data are available: 20 and 180 points, respectively. This data is used to construct the final MFSM. Choosing between Options A and B leads to different models.

\begin{figure}[!ht]
	\begin{center}
		\includegraphics[width=12cm]{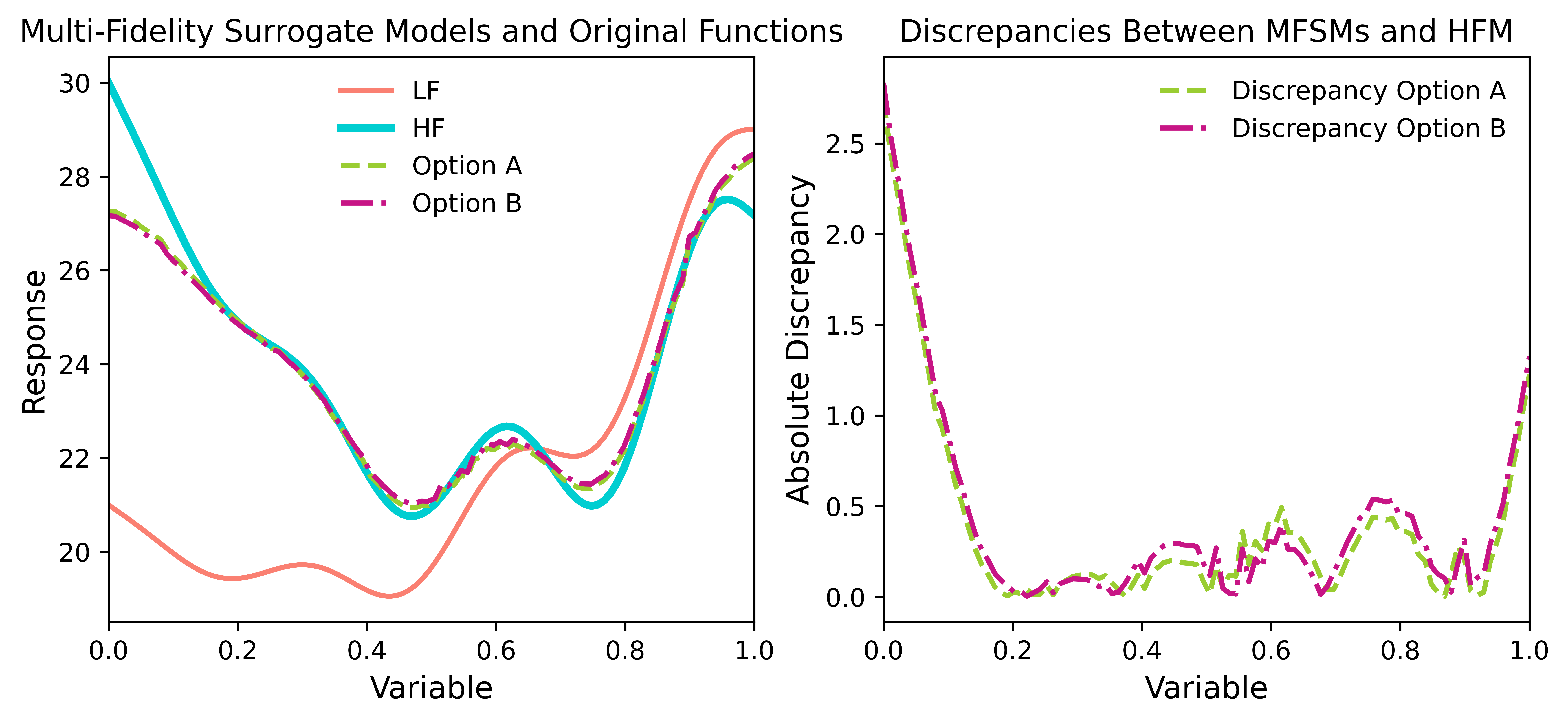}
		\caption{\label{fig:multchoice} One-dimensional analytic example illustrating the concept of comprehensive correction \cite{zhang2018multifidelity}.}
	\end{center}
\end{figure}

The reader interested in replicating these results is encouraged to access the example 3 \href{https://github.com/mgisellef/ReviewOfMultiFidelityModels_ToyProblems/blob/main/Multiple_Choice.ipynb}{notebook}.

\subsection{Example 4: Non-deterministic problem}

\begin{figure}[!ht]
	\begin{center}
		\includegraphics[width=8cm]{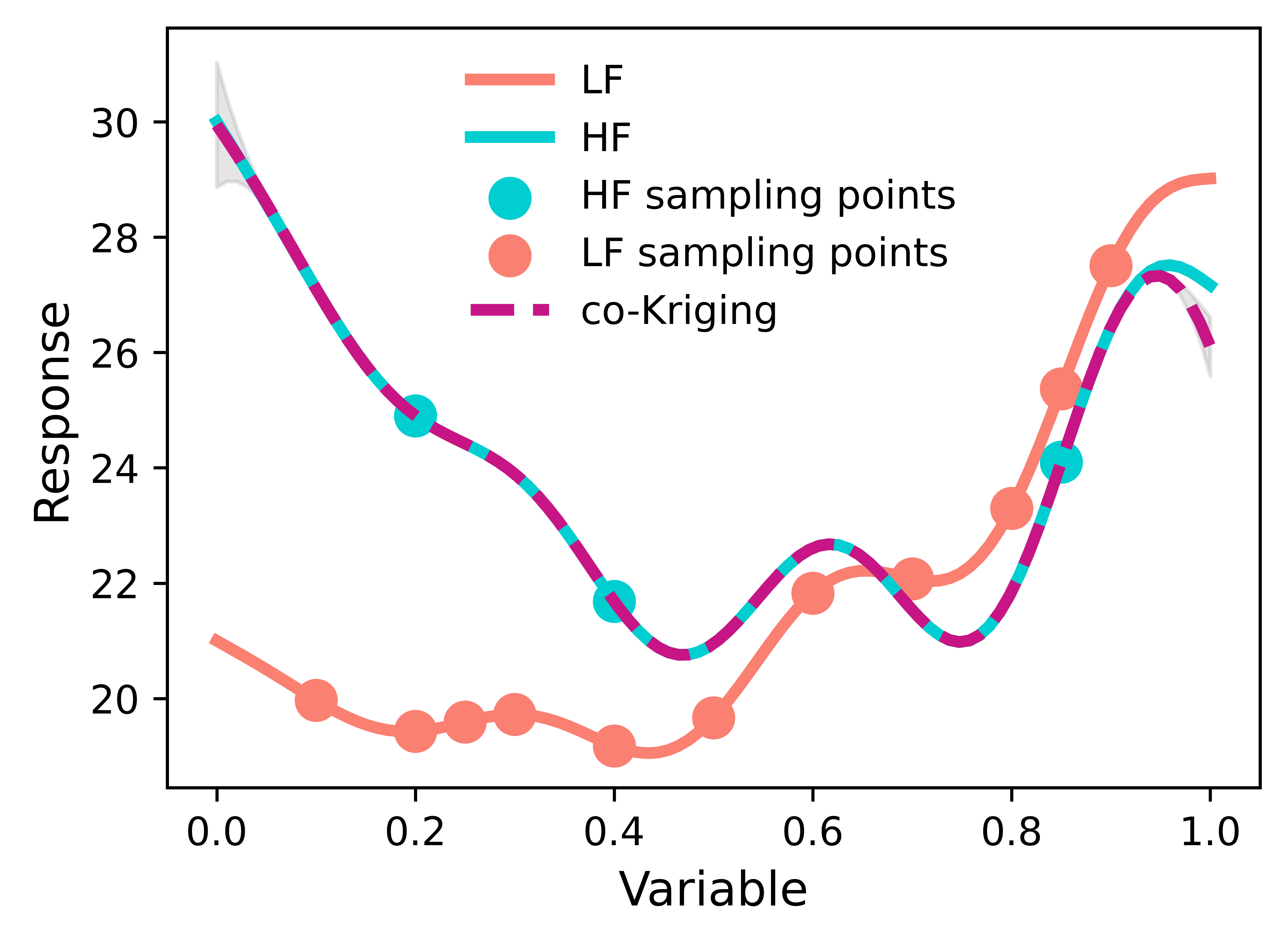}
		\caption{\label{fig:coKG} One-dimensional co-Kriging example. \cite{zhang2018multifidelity}.}
	\end{center}
\end{figure}

In this example, we explore a non-deterministic problem to exemplify the capabilities of co-Kriging in managing uncertainty. The problem involves modeling a system or process where outcomes are influenced by stochastic or random factors, introducing inherent variability in the results. MFSs are used to account for uncertainties and refine predictions effectively. In Figure \ref{fig:coKG}, the LF curve represents the LFM, and the HF curve represents the HFM. The scatter points depict the locations where high-fidelity evaluations were conducted. LF sampling points represent the positions where low-fidelity evaluations occurred. The dashed line labeled co-Kriging showcases the predicted response. The shaded region surrounding the co-Kriging prediction represents the confidence interval, providing insights into the uncertainty associated with the model's estimations. Figure \ref{fig:initial_sample} helps us understand the sensitivity of initial sampling for this co-Kriging example.

\begin{figure}[!ht]
	\begin{center}
		\includegraphics[width=10cm]{Initial_sample_sensitivity.jpg}
		\caption{\label{fig:initial_sample} Co-Kriging sensitivity to initial sampling points}
	\end{center}
\end{figure}

The reader interested in replicating these results is encouraged to access the  example 4 \href{https://github.com/mgisellef/ReviewOfMultiFidelityModels_ToyProblems/blob/main/Co-Kriging.ipynb}{notebook}.

\subsection{Example 5: Benchmark functions}

\begin{figure}[!ht]
	\begin{center}
		\includegraphics[width=8cm]{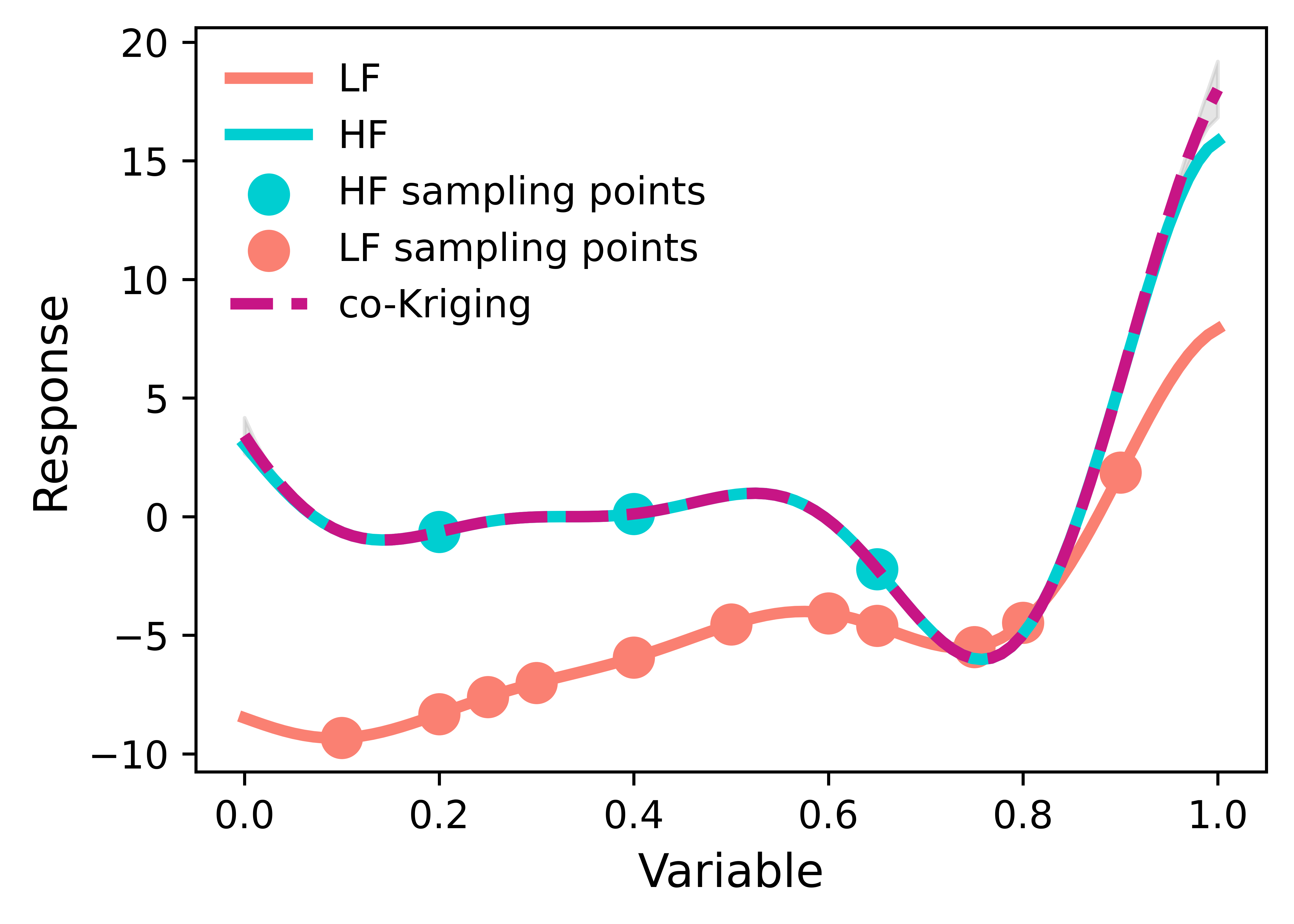}
		\caption{\label{fig:forrester} Predictive landscape of the Forrester function. Gray regions delineate areas falling within one standard deviation.}
	\end{center}
\end{figure}

The first function under consideration is the one-dimensional extended Forrester function ~\cite{forrester2008engineering}, whose HFM and LFM are depicted in Figure \ref{fig:forrester}. The co-Kriging approach was applied to build the MFM. The performance for the mean prediction is outstanding. The gray regions represent areas within one standard deviation from the mean, thereby providing an estimate of the associated uncertainty. As Figure \ref{fig:forrester} shows, the uncertainty estimation grows near the domain limits as expected. However, the MFM still under-predicts the variability of the function in those areas.

The next function considered is the two-dimensional multi-fidelity Branin (Eqs. \eqref{eq:B_HF} and \eqref{eq:B_LF}). The equation for the high-fidelity function is outlined as follows:
\begin{equation} \label{eq:B_HF}
	y_{HF}(x,y) = a \cdot \left( y - b \cdot x^2 + c \cdot x - r \right)^2 + s \cdot (1 - t) \cdot \cos(x) + s.
\end{equation}

In this particular implementation, the constants \( a \), \( b \), \( c \), \( r \), \( s \) and \( t \) are set as \( a = 1 \), \( b = \frac{5.1}{4\pi^2} \), \( c = \frac{5}{\pi} \), \( r = 6 \), \( s = 10 \) and \( t = \frac{1}{8\pi} \), respectively. The low-fidelity function is
\begin{equation} \label{eq:B_LF}
	y_{LF}(x,y) = 0.5 \cdot \text{func\_HF}(x, y) + 5 \cdot (y - 0.5) - 5.
\end{equation}

The function is showcased in Figure \ref{fig:branin}. Two MFM were created using additive and multiplicative corrections based on Eqs. \eqref{eq:B_HF} and \eqref{eq:B_LF}. The performance is assessed by the mean absolute percentage error (MAPE), which indicates that the accuracy of these approaches compares for this example.

\begin{figure}[!ht]
	\begin{center}
		\includegraphics[width=12cm]{3D_Branin_Function_Correction_plots.jpg}
		\caption{\label{fig:branin} Comparative performance of Branin function predictions employing both additive and multiplicative corrections. The associated MAPE is an evaluative metric, signaling comparable efficacy between the approaches.}
	\end{center}
\end{figure}

The reader interested in replicating these results is encouraged to access the  example 5 \href{https://github.com/mgisellef/ReviewOfMultiFidelityModels_ToyProblems/blob/main/Benchmark_Functions.ipynb}{notebook}.

\clearpage
\section{Tables}\phantom{text} \label{sec:tables}

\begin{table}[H]
	\begin{center}
		\caption{Categorization of fluid-mechanics-focused papers based on the methodologies employed as high- and low-fidelity models. The analysis techniques used in these studies were analyzed and categorized into six distinct categories: analytical approach (An), empirical methods (Em), linear analysis (Li), potential flow models (PF), Euler analysis (Eu) and Reynolds-averaged Navier-Stokes techniques (RANS).}
		\begin{tabular}{ >{\centering\arraybackslash}m{1.5in} | >{\centering\arraybackslash}m{0.4in} | >{\centering\arraybackslash}m{0.4in} | >{\centering\arraybackslash}m{0.3in} |>{\centering\arraybackslash}m{0.3in} |>{\centering\arraybackslash}m{0.3in}|>{\centering\arraybackslash}m{0.4in}}
			\multicolumn{7}{c}{\textbf{Fluid mechanics}} \\
			\hline
			\textbf{Reference} & \textbf{An} & \textbf{Em} & \textbf{Li} & \textbf{PF} & \textbf{Eu} & \textbf{RANS} \\
			\hline\noalign{\smallskip} \hline
			\cite{goldsmith2011effects} \cite{umakant2007ranking} &LF&- &HF&- &- & -\\ \hline
			\cite{allaire2014mathematical} \cite{bohnke2011integrated} \cite{burgee1996coarse} \cite{christensen2012multifidelity} \cite{forrester2006optimization} \cite{forrester2007multi}  &- &LF&HF&- &- &\\ \hline
			\cite{minisci2013robust} \cite{minisci2011robust} \cite{nguyen2013multidisciplinary} \cite{variyar2016multifidelity}& - & LF &- & -&- & HF \\ \hline
			\cite{burgee1994parallel} \cite{dufresne2008variable} \cite{geiselhart2011integration} \cite{knill1999response} \cite{kroo2010multifidelity} \cite{march2012constrained} \cite{march2012provably} \cite{padron2014multi} \cite{rajnarayan2008multifidelity} & - & - & LF & -& HF&- \\ \hline
			\cite{choi1997multi} \cite{deblois2010multi} \cite{toal2014multifidelity} \cite{zheng2015difference} \cite{zheng2013multi} & & & LF & & & HF\\ \hline
			\cite{bahrami2016multi} \cite{kandasamy2011multi} \cite{nelson2007multi} \cite{willis2008multifidelity} & - & - &- & LF & - & HF\\ \hline
			\cite{alexandrov2000first} \cite{ghoreyshi2008integration} \cite{han2012alternative} \cite{huang2013research} \cite{padron2016multi} \cite{ren2016multi}& - & - & &- & LF& HF\\ \hline
		\end{tabular}
		\label{tab:types_Fluids}
	\end{center}
\end{table}

\begin{table}[!ht]
	\begin{center}
		\caption{Distinct types of fidelity implemented in research papers on fluid mechanics differ from those based on analysis type. The classifications include dimensionality (2D/3D), analysis resolution (coarse vs. refined), type of study (simulations vs. experiments), state of flow (transient vs. steady) and degree of solution convergence (semiconverged vs. converged). The following abbreviations designate each paper's physical model: Em (empirical), Li (linear), PF (potential flow), Eu (Euler), RANS (Reynolds-averaged Navier-Stokes), URANS (unsteady RANS), TM (turbulence method), MHD (magnetohydrodynamics), AE (aeroelastic equations), MPF (multiphase flow) and TM (thermomechanical equations).}
		\begin{tabular}{ >{\centering\arraybackslash}m{2in} | >{\centering\arraybackslash}m{2.7in} }
			%
			\multicolumn{2}{c}{\textbf{Fluid mechanics}} \\
			\hline
			\textbf{Fidelity type} & \textbf{Reference} \\
			\hline\noalign{\smallskip} \hline
			\textbf{Dimensionality}& \cite{forrester2006optimization} 2D/3D Eu, \cite{joly2014integrated} 1D/3D RANS+TM, \cite{kandasamy2011multi} 2D/3D URANS, \cite{lazzara2009multifidelity} 2D/3D, \cite{qian2008bayesian} 1D/2D RANS, \cite{robinson2008surrogate} 1D/2D Li, \cite{turner2004multi} 1D/3D RANS, \cite{wang2013novel} 1D/3D RANS, \cite{zou2016shroud} 1D,2D/3D RANS\\ \hline
			\textbf{Coarse/Refined}& \cite{alexandrov2001approximation} Eu, \cite{brooks2011multi} RANS, \cite{choi2005multi} Eu, \cite{choi1997multi} Eu, \cite{fernandez2019linear} Eu, \cite{choi2008multifidelity} Li/Eu, \cite{jonsson2015shape} RANS, \cite{kennedy2000predicting} MPF, \cite{kleiber2013parameter} MHD, \cite{koziel2012knowledge} Eu, \cite{koziel2013multi}, Eu\cite{koziel2016rapid} Eu, \cite{lee2016efficiency} Eu, \cite{madsen2001multifidelity} RANS, \cite{rethore2011topfarm} RANS, \cite{toal2015some} RANS, \cite{zahir2013variable} Eu/RANS \\ \hline
			\textbf{Exp./Sim.}& \cite{fidkowski2014quantifying} Euler/MHD, \cite{forrester2010black} PF/Em, \cite{kuya2011multifidelity} RANS, \cite{toal2011efficient} RANS\\ \hline
			\textbf{Semiconverged/Converged}& \cite{jonsson2015shape}, RANS\cite{koziel2013multi} Eu\\ \hline
			\textbf{Steady/Transient}& \cite{berci2011multifidelity} AE, \cite{ghoman2012multifidelity} Eu, \cite{toal2015some} TM \\ \hline
		\end{tabular}
		\label{tab:extra_Fluids}
	\end{center}
\end{table}

\begin{table}[!ht]
	\begin{center}
		\caption{Categorization of papers within the domain of solid mechanics, according to the fidelity type employed in their analyses. The four fidelity types included are analytical (An), empirical (Em), linear (Li) and non-linear (NL).}
		\begin{tabular}{>{\centering\arraybackslash}m{1.5in}|>{\centering\arraybackslash}m{0.62in}|>{\centering\arraybackslash}m{0.62in}|>{\centering\arraybackslash}m{0.63in}|>{\centering\arraybackslash}m{0.63in}}
			%
			\multicolumn{5}{c}{\textbf{Solid mechanics}} \\
			\hline
			\textbf{Reference} & \textbf{An} & \textbf{Em} & \textbf{Li} & \textbf{NL}\\
			\hline\noalign{\smallskip} \hline
			\cite{sharma2008multi} & LF &- & HF &- \\ \hline
			\cite{venkataraman2006reliability} \cite{venkataraman1998design} &- & LF & HF &-\\ \hline
			\cite{venkataraman1998design} &- & LF &- & HF\\ \hline
			\cite{allaire2014multifidelity} \cite{allaire2010bayesian} \cite{drissaoui2012stochastic} \cite{hutchison1994variable} \cite{rajnarayan2008multifidelity} \cite{rodriguez2001sequential} \cite{viana2009optimization} \cite{vitali1997structural}&- & -& LF & HF\\ \hline
		\end{tabular}
		\label{tab:types_Solids}
	\end{center}
\end{table}

\begin{table}[ht]
	\begin{center}
		\caption{Categorization of solid mechanics research papers according to the type of fidelity used. The fidelity categories are defined based on dimensionality (i.e., 2D/3D), degree of refinement (i.e., coarse vs. refined) and simplification of boundary conditions (i.e., infinite plate vs. finite plate). Each paper is associated with the corresponding model employed, with Li indicating the use of linear models and NL representing non-linear models.}
		\begin{tabular}{ >{\centering\arraybackslash}m{1.5in} | >{\centering\arraybackslash}m{2.7in} }
			%
			\multicolumn{2}{c}{\textbf{Solid mechanics}} \\
			\hline
			\textbf{Fidelity type} & \textbf{Reference} \\
			\hline\noalign{\smallskip} \hline
			\textbf{Dimensionality} & \cite{le2015cokriging} 1D/2D Li, \cite{le2014bayesian} 1D/3D, \cite{mason1994analysis} 2D/3D, \cite{mason1998variable} 2D/3D Li, \cite{sharma2009multi} 2D/3D Li, \cite{singh2010mixed} 2D/3D NL \\ \hline
			\textbf{Coarse/Refined}&\cite{balabanov1998multifidelity} Li, \cite{biehler2015towards} NL, \cite{bohnke2011approach} Li, \cite{bradley2013multi} NL, \cite{chang1993sensitivity} Li, \cite{leary2003method} Li, \cite{march2011gradient} Li, \cite{sun2010two} NL, \cite{sun2011multi} NL, \cite{vitali1999correction} Li, \cite{zadeh2002multi} Li, \cite{zadeh2005use} Li\\ \hline
			\textbf{Boundary conditions}& \cite{vitali1998correction} Li, \cite{vitali2002structural} Li\\ \hline
		\end{tabular}
		\label{tab:extra_Solids}
	\end{center}
\end{table}

\begin{table}[!ht]
	\begin{center}
		\caption{Papers that use deterministic methods for constructing multi-fidelity surrogate models.}
		\begin{tabular}{ >{\centering\arraybackslash}m{1.5in} | >{\centering\arraybackslash}m{2.5in} }
			\multicolumn{2}{c}{\textbf{Deterministic methods}} \\
			\hline
			\textbf{Combining method} & \textbf{Reference} \\
			\hline\noalign{\smallskip} \hline
			\textbf{Additive correction} &\cite{balabanov1998multifidelity} \cite{balabanov1999reasonable} \cite{burgee1994parallel} \cite{eldred2009recent} \cite{goldfeld2005multi} \cite{knill1999response} \cite{padron2014multi} \cite{robinson2006strategies} \cite{robinson2006multifidelity} \cite{roux1998response} \cite{sharma2008multi} \cite{sharma2009multi} \cite{sun2011multi} \cite{sun2010two} \cite{toropov1998use} \cite{viana2009optimization} \cite{vitali2002structural}\\ \hline
			\textbf{Multiplicative correction} & \cite{alexandrov1998trust} \cite{alexandrov2001approximation}\cite{balabanov1998multifidelity} \cite{balabanov1999reasonable} \cite{burgee1996coarse} \cite{burgee1994parallel} \cite{chang1993sensitivity} \cite{goldfeld2005multi} \cite{haftka1991combining} \cite{hevesi1992precipitation} \cite{hutchison1994variable} \cite{kaufman1996variable} \cite{madsen2001multifidelity} \cite{mason1998variable} \cite{narayan2014stochastic} \cite{robinson2006strategies} \cite{robinson2006multifidelity} \cite{sharma2008multi} \cite{sharma2009multi} \cite{sun2011multi} \cite{sun2010two} \cite{toropov1998use} \cite{venkataraman2006reliability} \cite{venkataraman1998design} \cite{vitali1998correction} \cite{vitali1997structural} \cite{vitali2002structural} \cite{vitali1999correction} \\ \hline
			\textbf{Comprehensive correction} & \cite{eldred2006formulations} \cite{ghoreyshi2008integration} \cite{kim2007hybrid} \cite{ng2012multifidelity} \cite{roderick2014proper} \cite{zadeh2002multi} \cite{zadeh2005use}\\ \hline
			\textbf{Space mapping} & \cite{castro2006developing} \cite{jonsson2015shape} \cite{koziel2016rapid} \cite{ren2016multi} \cite{robinson2008surrogate} \\ \hline
		\end{tabular}
		\label{tab:DF}
	\end{center}
\end{table}

\begin{table}[!ht]
	\begin{center}
		\caption{Papers that use non-deterministic methods to construct multi-fidelity surrogate models.}
		\begin{tabular}{>{\centering\arraybackslash}m{1.5in} | >{\centering\arraybackslash}m{2.5in} }
			\multicolumn{2}{c}{\textbf{Non-deterministic methods}} \\
			\hline
			\textbf{Combining method} & \textbf{Reference} \\
			\hline\noalign{\smallskip} \hline
			\textbf{Additive correction} &\cite{biehler2015towards} \cite{forrester2006optimization} \cite{han2012alternative} \cite{kleiber2013parameter} \cite{march2012provably} \cite{march2011gradient} \cite{peherstorfer2016multifidelity} \cite{qian2008bayesian} \cite{rajnarayan2008multifidelity} \\ \hline
			\textbf{Multiplicative correction} &\cite{chen2015effective} \cite{forrester2010black} \cite{march2012constrained} \\ \hline
			\textbf{Comprehensive correction} & \cite{allaire2014mathematical} \cite{bradley2013multi} \cite{brooks2011multi}\cite{forrester2007multi} \cite{goh2013prediction} \cite{huang2013research} \cite{keane2012cokriging} \cite{kennedy2000predicting} \cite{kuya2011multifidelity} \cite{le2013multi} \cite{le2013bayesian} \cite{le2015cokriging} \cite{le2014bayesian} \cite{leary2003knowledge} \cite{perdikaris2015multi} \cite{toal2014potential} \cite{toal2011efficient} \cite{yu2014profile}\\ \hline
			\textbf{Calibration + comprehensive correction} &\cite{bayarri2007framework} \cite{fidkowski2014quantifying} \cite{higdon2004combining} \cite{kennedy2001bayesian} \\ \hline
		\end{tabular}
		\label{tab:NDF}
	\end{center}
\end{table}

\begin{table}[ht]
	\begin{center}
		\caption{MFM/HFM cost ratio. The references are divided also per field, given by fluid mechanics, solid mechanics and other. Other includes electronics, aeroelasticity, thermodynamics and analytical functions.}
		\begin{tabular}{>{\centering\arraybackslash}m{0.8in} | >{\centering\arraybackslash}m{1.0in} | >{\centering\arraybackslash}m{1.0in} | >{\centering\arraybackslash}m{1.0in}}
			\multicolumn{4}{c}{\textbf{MFM cost/HFM cost}}\\
			\hline
			\textbf{Percentage} & \textbf{Fluid mechanics} & \textbf{Solid mechanics} & \textbf{Other}\\
			\hline\noalign{\smallskip} \hline
			{0\% - 20\%} & \cite{choi2008multifidelity} \cite{padron2016multi} \cite{ren2016multi} \cite{ng2014multifidelity} & \cite{vitali2002multi} & \cite{kennedy2000predicting} \cite{koziel2010robust}  \cite{peherstorfer2016multifidelity} \cite{thoresson2007efficient} \\ \hline
			{21\% - 40\%} & \cite{alexandrov2000first} \cite{alexandrov2001approximation} \cite{padron2016multi} \cite{toal2014potential} \cite{toal2011efficient}  & \cite{burton2003variable} & - \\ \hline
			{41\% - 60\%} & \cite{alexandrov2001approximation} \cite{kleiber2013parameter} \cite{robinson2006strategies} \cite{robinson2006multifidelity} & \cite{qian2008bayesian} & \cite{jacobs2013computationally} \cite{molina2010multi}\\ \hline
			{61\% - 80\%}& \cite{jonsson2015shape} \cite{knill1999response} \cite{robinson2008surrogate}  & - & \cite{celik2010dddas} \cite{zahir2013variable} \cite{zou2016shroud}\\ \hline
			{81\% - 90\%}& - & \cite{balabanov1998multifidelity} & \cite{koyuncu2007dddas}\\ \hline
		\end{tabular}
		\label{tab:MF_HF}
	\end{center}
\end{table}


\medskip
Received May 2023; revised September 2023; early access November 2023.
\medskip

\end{document}